\documentclass[lettersize,journal]{IEEEtran}
\usepackage{graphicx}
\usepackage{cite}
\makeatletter
\let\NAT@parse\undefined
\makeatother
\usepackage{picinpar}
\usepackage{amsmath,amssymb,amsfonts,bm}
\usepackage{url}
\usepackage{flushend}
\usepackage[utf8]{inputenc}
\usepackage{colortbl}
\usepackage{soul}
\usepackage{multirow}
\usepackage{pifont}
\usepackage{xcolor}
\usepackage{alltt}
\usepackage{enumerate}
\usepackage{siunitx}
\usepackage{breakurl}
\usepackage{epstopdf}
\usepackage{pbox}
\usepackage{stfloats}
\usepackage{textcomp}
\usepackage{booktabs}
\usepackage{array}
\usepackage{placeins}
\usepackage{tikz}
\usepackage[hidelinks]{hyperref}

\graphicspath{{./}}

\hypersetup{
  pdftitle={Optimization-Based Velocity-Integral Sliding-Window Coarse Alignment: Attitude Error Analysis and Validation},
  pdfauthor={Xuyang Jiang, Xiyuan Chen, and Weiming Jing},
  pdfsubject={arXiv v3 preprint},
  pdfkeywords={attitude error covariance, coarse alignment, error propagation, optimization-based alignment, strapdown inertial navigation system}
}

\def\BibTeX{{\rm B\kern-.05em{\sc i\kern-.025em b}\kern-.08em
    T\kern-.1667em\lower.7ex\hbox{E}\kern-.125emX}}
\newcommand{\crossmat}[1]{\left[#1\times\right]}
\newcommand{\IEEEPreprintNotice}{%
  This work has been submitted to the IEEE for possible publication. Copyright may be transferred without notice, after which this version may no longer be accessible.%
}

\makeatletter
\let\arxiv@maketitle\maketitle
\let\arxiv@@maketitle\@maketitle
\let\arxiv@thanks\thanks
\makeatother

\begin{document}

\AddToHookNext{shipout/foreground}{%
  \begin{tikzpicture}[remember picture,overlay]
    \fill[white] (current page.north west) rectangle ([yshift=-42pt]current page.north east);
    \node[
      anchor=north,
      align=center,
      text width=0.94\paperwidth,
      font=\sffamily\fontsize{7}{8.2}\selectfont
    ] at ([yshift=-8pt]current page.north) {\IEEEPreprintNotice};
  \end{tikzpicture}%
}

\title{Optimization-Based Velocity-Integral Sliding-Window Coarse Alignment: Attitude Error Analysis and Validation}

\markboth{IEEE/ASME TRANSACTIONS ON MECHATRONICS}
{Jiang: Error Analysis of Sliding-Window OBA Coarse Alignment}

\author{Xuyang Jiang, \IEEEmembership{Graduate Student Member, IEEE},
Xiyuan Chen, \IEEEmembership{Senior Member, IEEE},
and Weiming Jing, \IEEEmembership{Graduate Student Member, IEEE}
\thanks{This work was supported in part by Guizhou Provincial Key Technology R\&D Program under Grant XKBF [2025] 032, and in part by the National Natural Science Foundation of China under Grant 61873064. (Corresponding author: Xiyuan Chen.)}
\thanks{Xuyang Jiang and Weiming Jing are with the School of Instrument Science and Engineering, State Key Laboratory of comprehensive PNT Network and Equipment Technology, Southeast University, Nanjing 210096, China, and also with the Key Laboratory of MicroInertial Instrument and Advanced Navigation Technology of Ministry of Education, Southeast University, Nanjing 210096, China (e-mail: jxy@seu.edu.cn; jwm@seu.edu.cn).}
\thanks{Xiyuan Chen is with the School of Instrument Science and Engineering, State Key Laboratory of comprehensive PNT Network and Equipment Technology, Southeast University, Nanjing 210096, China, also with the Key Laboratory of Micro-Inertial Instrument and Advanced Navigation Technology of Ministry of Education, Southeast University, Nanjing 210096, China, and also with the School of Aerospace Engineering, Guizhou Institute of Technology, Guizhou 550025, China (e-mail: chxiyuan@seu.edu.cn).}}

\maketitle

\begin{abstract}
The optimization-based alignment (OBA) approach transforms strapdown inertial navigation system (SINS) coarse alignment into a constant initial attitude estimation problem for global navigation satellite system (GNSS)-aided in-motion alignment. While existing studies mainly improve accuracy by refining attitude determination algorithms or constructing robust observation vectors, a rigorous analytical mapping from raw sensor and aiding-velocity uncertainties to attitude errors remains unavailable for fixed-length sliding-window velocity-integral OBA. To address this issue, this paper proposes a first-order attitude error propagation model. Based on the sliding-window observation model, gyroscope errors, accelerometer errors, GNSS velocity noise, and lever-arm effects are propagated to unnormalized observation-vector perturbations, which are further mapped to attitude misalignment through Davenport's q method. The model decouples systematic errors from stochastic noise and derives the corresponding deterministic attitude offsets and error covariances. Monte Carlo simulations demonstrate that the analytical model captures deterministic offsets and statistical spread, yielding standard-deviation ratios between 0.942 and 1.036 with empirical coverage above 99.4\%. Vehicle field tests show that the predicted covariance envelopes bound the actual initial-attitude errors, with the maximum residual root-mean-square error (RMSE) below 0.00495 deg. These results validate the proposed model for coarse-alignment attitude error assessment.
\end{abstract}


\section{Introduction}
\IEEEPARstart{S}{trapdown} inertial navigation system (SINS) propagates attitude, velocity, and position without continuous external measurements and is fundamental to terrestrial, surface-vessel, airborne, and underwater navigation systems \cite{GeLuFuGNSSYuGuanXingJiDuoChuanGanQiZuHeDaoHangXiTongYuanLi2015}. As a dead-reckoning system, a SINS must be initialized with reliable navigation states. Velocity and position are typically provided by external sensors such as the global navigation satellite system (GNSS), Doppler velocity log (DVL), or odometer, whereas the initial attitude must be determined by alignment. Fine alignment usually relies on an extended Kalman filter (EKF), whose linearized error model is valid only when the attitude misalignment is sufficiently small. Although nonlinear or data-driven filtering schemes can handle large initial misalignments, they usually require more computation and more delicate parameter tuning \cite{liuStrongTrackingUKFBased2023,xuDualFreeSizeLSSVM2024}. Therefore, accurate and robust coarse alignment remains indispensable. Beyond merely estimating the initial attitude, practical implementation also requires a reliable error covariance to initialize subsequent fine-alignment filters \cite{frutuosoPerformanceEvaluationCoarse2023}, making explicit error evaluation as important as the alignment algorithm itself.

SINS coarse alignment can be interpreted as a multi-vector attitude determination problem. In redundant vector-observation cases, Wahba's problem formulates attitude determination as a least-squares optimization over attitude matrices, with singular value decomposition (SVD), fast optimal attitude matrix (FOAM), Davenport's q method, quaternion estimator (QUEST), and estimator of the optimal quaternion (ESOQ) as representative solution strategies \cite{markleyFundamentalsSpacecraftAttitude2014}. To adapt this multi-vector attitude-determination framework to coarse alignment, early studies constructed observation vectors from the apparent drift of gravity in the inertial frame or from aided velocity measurements \cite{kinseyAdaptiveIdentificationGroup2007,guCoarseAlignmentMarine2008}. Optimization-based alignment (OBA) and velocity-loci alignment then showed that accumulated apparent velocity vectors can transform in-motion coarse alignment into a constant initial attitude estimation problem \cite{wuOptimizationbasedAlignmentInertial2011,silsonCoarseAlignmentShips2011}. Compared with analytical coarse alignment, this accumulation improves redundancy and strengthens maneuver-dependent observability. Wu further analyzed the global observability of SINS alignment under suitable maneuvers from a nonlinear constraint perspective \cite{wuObservabilityStrapdownINS2012}. Building on this OBA framework, velocity-integration OBA further incorporated standard SINS coning and sculling compensation into the discrete velocity-integral formulation for GNSS-aided in-motion coarse alignment \cite{wuVelocityPositionIntegration2013a}.

Subsequent studies have improved OBA along several complementary directions. For swaying bases, dynamic sliding-window strategies and equivalent backtracking have been used to suppress observation-vector errors \cite{yueEquivalentBacktrackingCoarse2025}. Wu, Chang, and Huang incorporated gyroscope bias, accelerometer bias, lever-arm parameters, and body-attitude propagation errors into online optimization, nonlinear filtering, or EKF closed-loop feedback frameworks for GNSS-aided and odometer-aided SINS coarse alignment \cite{wuNewTechniqueINS2014,changOptimizationbasedAlignmentStrapdown2016,huangKalmanFilteringBasedInMotionCoarse2017,huangNewFastInMotion2018a,huangHighAccuracyGPSAidedCoarse2020}. More recent work has extended the formulation itself through kinematic constraints \cite{xuRobustInMotionAlignment2021,qinFastInMotionAlignment2024,liDVLAidedInMotionCoarse2023,yaoSINSInMotionCoarse2025}, vector subtraction, magnitude matching, robust weighting \cite{xuRobustInMotionOptimizationBased2022,zhouRobustInMotionCoarse2025}, sliding-window optimization \cite{xuRobustAttitudePositioning2024}, filtering based on the vectorized Davenport gain matrix \cite{huangCoarseAlignmentMethod2023}, and factor graph optimization (FGO) \cite{zhouUnifiedInitialAlignment2023}, as well as by employing Lie-group invariant-error descriptions for complex dynamics and large-misalignment conditions \cite{changLogLinearErrorState2023,tangInvariantErrorBasedIntegrated2023a}. Taken together, these studies indicate that the attitude error of sliding-window velocity-integral OBA is jointly determined by deterministic sensor errors, stochastic sensor noise, aiding-velocity errors, lever-arm effects, and motion excitation. Consequently, providing only an attitude estimate without quantifying its uncertainty is insufficient for assessing alignment reliability.

Despite these algorithmic advancements, error propagation analysis for OBA remains insufficiently addressed. For general attitude determination, Shuster proposed the TASTE test for unit-vector observation consistency \cite{shusterTASTETest2009}, while Markley and Crassidis analyzed attitude-estimation covariance for Wahba's problem using formulations based on QUEST and Davenport's q method under unit-vector observation and zero-mean Gaussian error assumptions \cite{markleyFundamentalsSpacecraftAttitude2014}. For SINS alignment, Silson derived covariance expressions for stationary velocity-loci alignment, while Wu developed a linear error model for gyroscope bias, accelerometer bias, and lever-arm effects \cite{silsonCoarseAlignmentShips2011,wuNewTechniqueINS2014}. Within analytical coarse alignment, Silva further derived stationary SINS error equations and bias-estimation formulations \cite{silvaErrorAnalysisAnalytical2016,silvaFastInFieldCoarse2018}. For unnormalized observations, Chang analyzed Davenport's q method error propagation, Cheng extended attitude estimation to pose-estimation covariance analysis, and Ouyang and Wu developed covariance analyses for stationary, odometer-aided, and GNSS-aided OBA scenarios \cite{changErrorAnalysisDavenports2017,chengOptimalPoseEstimation2021,ouyangOptimizationBasedStrapdownAttitude2022}. Nonetheless, these derivations either rely on normalized unit-vector assumptions, focus on stationary or weakly dynamic cases, or require raw measurement errors to be independent and zero-mean Gaussian. Consequently, sliding-window velocity-integral OBA still lacks an explicit description of how deterministic attitude offsets, stochastic error covariance, accumulated attitude propagation errors, and window endpoint differencing are coupled under complex in-motion conditions.

To address this issue, this paper develops a first-order analytical propagation model for mapping raw measurement errors to attitude-error statistics for GNSS-aided sliding-window velocity-integral OBA. The main contributions are summarized as follows:
\begin{enumerate}[1)]
\item A compact sliding-window velocity-integral OBA observation model and its discrete implementation are formulated. Unlike initial-epoch integration, the fixed-window construction limits accelerometer-error accumulation but retains the accumulated effects of gyroscope errors in the attitude-chain projection, thereby providing the basis for the subsequent error analysis.

\item A first-order analytical propagation model from raw measurement errors to attitude misalignment is derived. Gyroscope errors, accelerometer errors, GNSS velocity errors, and lever-arm effects are mapped to observation-vector perturbations and Davenport gain matrix perturbations.

\item Deterministic and stochastic attitude-error components are analytically separated. The deterministic component predicts systematic attitude offsets, whereas the stochastic component incorporates temporal correlations caused by overlapping windows and yields the attitude-error covariance. Both the predicted offsets and covariance envelopes are validated by Monte Carlo simulations and vehicle field tests.
\end{enumerate}

The remainder of this paper is organized as follows. Section II presents the sliding-window velocity-integral OBA model, its discretization, and the Wahba solution with Davenport's q method. Section III derives the error propagation model. Section IV verifies the model by simulation. Section V presents a vehicle field-test validation, and Section VI concludes the paper.

\section{Sliding-Window Velocity-Integral OBA}
\subsection{Continuous Observation Model}
The east-north-up (ENU) navigation frame is denoted by $n$, and the right-forward-up (RFU) body frame by $b$. With the initial alignment epoch $t=0$ as the reference, the attitude matrix at any time $t$ is decomposed as
\begin{equation}
\mathbf{C}_b^n(t)=\mathbf{C}_{n(0)}^{n(t)} \mathbf{C}_b^n(0) \mathbf{C}_{b(t)}^{b(0)} ,
\label{eq:chain}
\end{equation}
where $\mathbf{C}_b^n(0)$ is the constant unknown initial attitude matrix. The matrices $\mathbf{C}_{b(t)}^{b(0)}$ and $\mathbf{C}_{n(t)}^{n(0)}$ describe the attitude changes of the body and navigation frames over $[0,t]$, respectively; the matrix $\mathbf{C}_{n(0)}^{n(t)}$ in \eqref{eq:chain} is the inverse of $\mathbf{C}_{n(t)}^{n(0)}$. These attitude-chain matrices are obtained by integrating the gyroscope-measured body angular rate $\boldsymbol{\omega}_{ib}^b$ and the velocity-derived navigation angular rate $\boldsymbol{\omega}_{in}^n$:
\begin{equation}
\dot{\mathbf{C}}_{b(t)}^{b(0)}
=\mathbf{C}_{b(t)}^{b(0)}\crossmat{\boldsymbol{\omega}_{ib}^b},
\quad
\dot{\mathbf{C}}_{n(t)}^{n(0)}
=\mathbf{C}_{n(t)}^{n(0)}\crossmat{\boldsymbol{\omega}_{in}^n},
\label{eq:frame_prop_cont}
\end{equation}
with \(\boldsymbol{\omega}_{in}^{n}=\boldsymbol{\omega}_{ie}^{n}+\boldsymbol{\omega}_{en}^{n}\), where \(\boldsymbol{\omega}_{ie}^{n}\) and \(\boldsymbol{\omega}_{en}^{n}\) denote the Earth rotation rate and transport rate, respectively, and \(\crossmat{\cdot}\) denotes the skew-symmetric cross-product matrix. The strapdown inertial specific-force equation is
\begin{equation}
\dot{\boldsymbol{v}}^n
=\mathbf{C}_b^n\boldsymbol{f}^b
-\left(2\boldsymbol{\omega}_{ie}^n+\boldsymbol{\omega}_{en}^n\right)
\times\boldsymbol{v}^n+\boldsymbol{g}^n .
\label{eq:velocity_equation}
\end{equation}
Here $\boldsymbol{v}^n=[v_E,v_N,v_U]^{\mathrm{T}}$ is the navigation-frame velocity, $\boldsymbol{f}^b$ is the specific force, and $\boldsymbol{g}^n$ is gravity. Substituting \eqref{eq:chain} into \eqref{eq:velocity_equation} and integrating over the sliding window \([t_m,t]\) yield the observation equation
\begin{equation}
\mathbf{C}_b^n(0)\boldsymbol{\alpha}(t)=\boldsymbol{\beta}(t),
\label{eq:obs_relation}
\end{equation}
where the body-side observation vector is
\begin{equation}
\boldsymbol{\alpha}(t)
=\int_{t_m}^{t} \mathbf{C}_{b(\tau)}^{b(0)}\boldsymbol{f}^b(\tau)d\tau ,
\label{eq:alpha_cont}
\end{equation}
and the navigation-side reference vector is
\begin{equation}
\begin{aligned}
\boldsymbol{\beta}(t)
&=\mathbf{C}_{n(t)}^{n(0)}\boldsymbol{v}^n(t)
-\mathbf{C}_{n(t_m)}^{n(0)}\boldsymbol{v}^n(t_m)\\
&\quad+
\int_{t_m}^{t}\mathbf{C}_{n(\tau)}^{n(0)}
\left[
\boldsymbol{\omega}_{ie}^n(\tau)\times\boldsymbol{v}^n(\tau)
-\boldsymbol{g}^n(\tau)
\right]d\tau .
\end{aligned}
\label{eq:beta_cont}
\end{equation}
Equations \eqref{eq:obs_relation}--\eqref{eq:beta_cont} follow the velocity-integral formulation of OBA \cite{wuOptimizationbasedAlignmentInertial2011,wuVelocityPositionIntegration2013a}. The key distinction is that each observation vector is constructed in a fixed-length sliding window rather than by an integral starting from the initial alignment time. This limits the integration length of a single observation vector and suppresses the long-term accumulation of accelerometer errors. Nevertheless, the attitude-chain matrix $\mathbf{C}_{b(\tau)}^{b(0)}$ appearing in the window integral retains the gyroscope errors accumulated prior to the window start.

\subsection{Discrete Implementation}
Let $t_k=kT$ be the discrete sampling epoch, where $T$ is the update period, and let the $i$th fixed-length sliding window be $[t_{m_i},t_{M_i}]$. Within one update period, the dual-sample gyroscope angle increments and accelerometer velocity increments are denoted by $\Delta\boldsymbol{\theta}_{1,k}$, $\Delta\boldsymbol{\theta}_{2,k}$, $\Delta\boldsymbol{v}_{1,k}$, and $\Delta\boldsymbol{v}_{2,k}$. The body-frame equivalent rotation vector with coning compensation is
\begin{equation}
\boldsymbol{\phi}_{b,k}
=\Delta\boldsymbol{\theta}_{1,k}
+\Delta\boldsymbol{\theta}_{2,k}
+\frac{2}{3}\Delta\boldsymbol{\theta}_{1,k}
\times\Delta\boldsymbol{\theta}_{2,k}.
\label{eq:coning}
\end{equation}
With dual-sample rotation and sculling compensation, the specific-force increment is
\begin{equation}
\begin{split}
\Delta\boldsymbol{v}_{c,k}
&=\Delta\boldsymbol{v}_{1,k}+\Delta\boldsymbol{v}_{2,k}\\
&\quad+\frac{1}{2}
(\Delta\boldsymbol{\theta}_{1,k}+\Delta\boldsymbol{\theta}_{2,k})
\times
(\Delta\boldsymbol{v}_{1,k}+\Delta\boldsymbol{v}_{2,k})\\
&\quad+\frac{2}{3}
(\Delta\boldsymbol{\theta}_{1,k}\times\Delta\boldsymbol{v}_{2,k}
+\Delta\boldsymbol{v}_{1,k}\times\Delta\boldsymbol{\theta}_{2,k}).
\end{split}
\label{eq:sculling}
\end{equation}
The discrete observation vectors of the $i$th sliding window are
\begin{equation}
\boldsymbol{\alpha}_i
=\sum_{k=m_i}^{M_i-1}\mathbf{C}_{b(t_k)}^{b(0)}\Delta\boldsymbol{v}_{c,k},
\label{eq:alpha_disc}
\end{equation}
and
\begin{equation}
\begin{aligned}
\boldsymbol{\beta}_i
&=\mathbf{C}_{n(t_{M_i})}^{n(0)}\boldsymbol{v}^n(t_{M_i})
-\mathbf{C}_{n(t_{m_i})}^{n(0)}\boldsymbol{v}^n(t_{m_i})\\
&\quad+
\sum_{k=m_i}^{M_i-1}\mathbf{C}_{n(t_k)}^{n(0)}
\left[
\boldsymbol{\omega}_{ie}^n(t_k)\times\boldsymbol{v}^n(t_k)
-\boldsymbol{g}^n(t_k)
\right]T .
\end{aligned}
\label{eq:beta_disc}
\end{equation}
The derivations of \(\boldsymbol{\alpha}(t)\), \(\boldsymbol{\beta}(t)\), and of \(\Delta\boldsymbol{v}_{c,k}\), \(\mathbf{C}_{b(t_k)}^{b(0)}\), and \(\mathbf{C}_{n(t_k)}^{n(0)}\), are provided in Supplementary Material, Secs.~I and II, respectively.

\subsection{Davenport's q Method Solution of the Wahba Problem}
Given $N$ sliding-window observation pairs $\{(\boldsymbol{\alpha}_i,\boldsymbol{\beta}_i)\}_{i=1}^{N}$, the weighted Wahba problem over the special orthogonal group $SO(3)$ is
\begin{equation}
\hat{\mathbf{C}}
=\arg\min_{\mathbf{C}\in SO(3)}
\sum_{i=1}^{N}w_i
\left\|
\boldsymbol{\beta}_i-\mathbf{C}\boldsymbol{\alpha}_i
\right\|^2 .
\label{eq:wahba}
\end{equation}
This paper uses the unnormalized velocity-integral observation vectors \(\boldsymbol{\alpha}_i\) and \(\boldsymbol{\beta}_i\) in the Wahba problem. Their magnitudes therefore reflect the local motion excitation within each sliding window, implicitly assigning higher weights to more informative observation vectors. The nominal implementation adopts \(w_i=1\), since any common scalar weight applied to all windows only scales the objective function and yields the same optimal attitude solution.

Define the attitude profile matrix $\mathbf{B}$, antisymmetric vector $\boldsymbol{z}$, and Davenport gain matrix $\mathbf{K}$ as
\begin{equation}
\mathbf{B}=\sum_{i=1}^{N}w_i\boldsymbol{\beta}_i\boldsymbol{\alpha}_i^{\mathrm{T}},
\quad
\boldsymbol{z}=\sum_{i=1}^{N}w_i\boldsymbol{\beta}_i\times\boldsymbol{\alpha}_i ,
\label{eq:BZ}
\end{equation}
\begin{equation}
\mathbf{K}=
\begin{bmatrix}
\operatorname{tr}(\mathbf{B}) & \boldsymbol{z}^{\mathrm{T}}\\
\boldsymbol{z} & \mathbf{B}+\mathbf{B}^{\mathrm{T}}-\operatorname{tr}(\mathbf{B})\mathbf{I}_3
\end{bmatrix}.
\label{eq:K}
\end{equation}
Using a scalar-first unit quaternion $\boldsymbol{q}=[s,\boldsymbol{\eta}^{\mathrm{T}}]^{\mathrm{T}}$, \eqref{eq:wahba} is equivalent to
\begin{equation}
\hat{\boldsymbol{q}}=\arg\max_{\boldsymbol{q}^{\mathrm{T}}\boldsymbol{q}=1}\boldsymbol{q}^{\mathrm{T}}\mathbf{K}\boldsymbol{q} .
\label{eq:qmax}
\end{equation}
Therefore, $\hat{\boldsymbol{q}}$ is the unit eigenvector associated with the largest eigenvalue of $\mathbf{K}$. Following the residual-minimization OBA formulation in~\cite{ouyangOptimizationBasedStrapdownAttitude2022}, the corresponding symmetric quadratic-form matrix satisfies $\mathbf{K}_{\mathrm{OBA}}=c\mathbf{I}_4-2\mathbf{K}$, where $c=\sum_{i=1}^{N}w_i(\|\boldsymbol{\alpha}_i\|^2+\|\boldsymbol{\beta}_i\|^2)$.
Therefore, the largest-eigenvalue eigenvector of \(\mathbf{K}\) is equivalent to the smallest-eigenvalue eigenvector of \(\mathbf{K}_{\mathrm{OBA}}\). A proof of this eigenvalue equivalence is provided in Supplementary Material, Sec.~III.

\section{Error Propagation Model}
\subsection{From Raw Errors to Observation-Vector Errors}
This subsection maps gyroscope errors, accelerometer errors, GNSS velocity errors, and lever-arm effects into equivalent perturbations of the observation vectors \(\boldsymbol{\alpha}_i\) and \(\boldsymbol{\beta}_i\).
Following the convention that the measurement error is defined as the measured value minus the true value, the gyroscope, accelerometer, and GNSS velocity measurement models are
\begin{equation}
\begin{aligned}
\widetilde{\boldsymbol{\omega}}_{ib}^{b}&=\boldsymbol{\omega}_{ib}^{b}+\delta\boldsymbol{\omega}_{ib}^{b},&
\delta\boldsymbol{\omega}_{ib}^{b}&=\boldsymbol{\epsilon}^{b}+\boldsymbol{w}_g,\\
\widetilde{\boldsymbol{f}}^{b}&=\boldsymbol{f}^{b}+\delta\boldsymbol{f}^{b},&
\delta\boldsymbol{f}^{b}&=\boldsymbol{\nabla}^{b}+\boldsymbol{w}_a,\\
\widetilde{\boldsymbol{v}}_{GNSS}^{n}&=\boldsymbol{v}_{GNSS}^{n}+\delta\boldsymbol{v}_{GNSS}^{n},&
\delta\boldsymbol{v}_{GNSS}^{n}&=\boldsymbol{w}_v.
\end{aligned}
\label{eq:meas_model}
\end{equation}
Here, \(\boldsymbol{\epsilon}^{b}\) and \(\boldsymbol{\nabla}^{b}\) are constant gyroscope and accelerometer biases, respectively, whereas \(\boldsymbol{w}_g\), \(\boldsymbol{w}_a\), and \(\boldsymbol{w}_v\) are modeled as mutually independent, zero-mean Gaussian white-noise processes. If $\boldsymbol{l}^b$ denotes the lever arm from the inertial measurement unit (IMU) measurement center to the GNSS antenna phase center, the GNSS antenna velocity and the IMU-center velocity satisfy
\begin{equation}
\begin{aligned}
\tilde{\boldsymbol{v}}_{GNSS}^n
&=\boldsymbol{v}^n
+\delta\boldsymbol{v}_{GNSS}^n
+\mathbf{C}_b^n(\boldsymbol{\omega}_{eb}^b\times\boldsymbol{l}^b),\\
\boldsymbol{\omega}_{eb}^b
&=\boldsymbol{\omega}_{ib}^b-\mathbf{C}_n^b\boldsymbol{\omega}_{ie}^n .
\end{aligned}
\label{eq:lever_velocity}
\end{equation}

To model the accumulated projection errors in \(\mathbf{C}_{b(\tau)}^{b(0)}\) and \(\mathbf{C}_{n(\tau)}^{n(0)}\), the computed attitude-chain matrix \(\widetilde{\mathbf{C}}_{x(\tau)}^{x(0)}\) for \(x\in\{b,n\}\) is expressed as the true matrix \(\mathbf{C}_{x(\tau)}^{x(0)}\) right-multiplied by a small-misalignment perturbation \(\boldsymbol{\phi}^x\):
\begin{equation}
\widetilde{\mathbf{C}}_{x(\tau)}^{x(0)}
\approx
\mathbf{C}_{x(\tau)}^{x(0)}
\left(\mathbf{I}_3+\crossmat{\boldsymbol{\phi}^x(\tau)}\right).
\label{eq:right_error}
\end{equation}
Under the first-order projection-error model, the accumulated misalignment \(\boldsymbol{\phi}^x\) induced by the angular-rate error \(\delta\boldsymbol{\omega}^x\) is expressed as
\begin{equation}
\boldsymbol{\phi}^x(\tau)
=
\int_0^\tau
\mathbf{C}_{x(s)}^{x(\tau)}
\delta\boldsymbol{\omega}^x(s)\,ds .
\label{eq:phi_dyn}
\end{equation}
The detailed derivation of \eqref{eq:phi_dyn} is provided in Supplementary Material, Sec.~IV.
For the body side, set \(x=b\) and \(\delta\boldsymbol{\omega}^x=\delta\boldsymbol{\omega}_{ib}^b\). Taking the first-order variation of \eqref{eq:alpha_cont} and using \eqref{eq:right_error} separates the direct accelerometer-error contribution from the gyroscope-induced projection error:
\begin{equation}
\delta\boldsymbol{\alpha}_{a,i}
=\int_{t_{m_i}}^{t_{M_i}}
\mathbf{C}_{b(\tau)}^{b(0)}
\delta\boldsymbol{f}^b(\tau)\,d\tau .
\label{eq:alpha_accel}
\end{equation}
\begin{equation}
\delta\boldsymbol{\alpha}_{g,i}
=-\int_{t_{m_i}}^{t_{M_i}}
\mathbf{C}_{b(\tau)}^{b(0)}
\crossmat{\boldsymbol{f}^b(\tau)}
\boldsymbol{\phi}^b(\tau)\,d\tau .
\label{eq:alpha_gyro}
\end{equation}
Equation \eqref{eq:alpha_gyro} reveals advantages of the sliding-window scheme. In conventional OBA, integrating from the initial alignment time causes late-stage observation vectors to grow continuously in magnitude. Since the unweighted Wahba problem implicitly favors vectors with larger magnitudes, observation pairs formed at later epochs of the alignment interval may exert greater influence on the attitude solution while being affected by larger accumulated gyroscope errors. By restricting the integration to a fixed length, the sliding-window approach bounds the vector magnitudes to reflect local motion intensity rather than the accumulated alignment time. This prevents heavily corrupted late-stage observations from disproportionately dominating the unweighted objective function.

For the navigation side, set \(x=n\) and \(\delta\boldsymbol{\omega}^x=\delta\boldsymbol{\omega}_{in}^n\). Under the adopted error model, the Earth-rotation rate is treated as known and position errors are neglected; therefore, the perturbation of \(\boldsymbol{\omega}_{in}^{n}\) is attributed to the velocity-dependent transport rate \(\boldsymbol{\omega}_{en}^{n}\), yielding \(\delta\boldsymbol{\omega}_{in}^{n}\approx\delta\boldsymbol{\omega}_{en}^{n}\). Taking the first-order variation of \eqref{eq:beta_cont} and using \eqref{eq:right_error} to separate the projection-matrix perturbation in \(\mathbf{C}_{n(\tau)}^{n(0)}\) from the perturbations introduced directly by \(\delta\boldsymbol{v}_{GNSS}^n\), the navigation-side observation-vector error consists of the endpoint velocity perturbation, the integral term associated with \(\boldsymbol{\omega}_{ie}^n\times\delta\boldsymbol{v}_{GNSS}^n\), and the accumulated projection perturbation induced by \(\boldsymbol{\phi}^n\). In the fixed-length sliding-window formulation, the latter two terms are scaled by the Earth's rotation rate and the radii of curvature and are therefore treated as secondary relative to the endpoint velocity perturbation. Retaining the dominant endpoint term gives
\begin{equation}
\delta\boldsymbol{\beta}_{v,i}
\approx
\mathbf{C}_{n(t_{M_i})}^{n(0)}\delta\boldsymbol{v}_{GNSS}^n(t_{M_i})
-\mathbf{C}_{n(t_{m_i})}^{n(0)}\delta\boldsymbol{v}_{GNSS}^n(t_{m_i}) .
\label{eq:beta_v}
\end{equation}

The lever-arm effect enters the navigation-side observation vector \(\boldsymbol{\beta}_i\) through the GNSS velocity measurement. Retaining the dominant contribution of \(\mathbf{C}_b^n(\boldsymbol{\omega}_{eb}^b\times\boldsymbol{l}^b)\) in \(\boldsymbol{\beta}_i\) and using the attitude-chain identity in \eqref{eq:chain} yield the following equivalent correction on the \(\boldsymbol{\alpha}_i\) side:
\begin{equation}
\begin{aligned}
\delta\boldsymbol{\alpha}_{l,i}
&\approx
\mathbf{C}_{b(t_{M_i})}^{b(0)}
\left[\boldsymbol{\omega}_{eb}^b(t_{M_i})\times\boldsymbol{l}^b\right]\\
&\quad-
\mathbf{C}_{b(t_{m_i})}^{b(0)}
\left[\boldsymbol{\omega}_{eb}^b(t_{m_i})\times\boldsymbol{l}^b\right].
\end{aligned}
\label{eq:alpha_lever}
\end{equation}
The equivalent observation-vector errors entering Davenport's q method are
\begin{equation}
\delta\boldsymbol{\alpha}_i
=\delta\boldsymbol{\alpha}_{a,i}
+\delta\boldsymbol{\alpha}_{g,i}
-\delta\boldsymbol{\alpha}_{l,i},
\quad
\delta\boldsymbol{\beta}_i
=\delta\boldsymbol{\beta}_{v,i}.
\label{eq:obs_error}
\end{equation}
The derivations of \(\delta\boldsymbol{\alpha}_{a,i}\), \(\delta\boldsymbol{\alpha}_{g,i}\), \(\delta\boldsymbol{\beta}_{v,i}\), and \(\delta\boldsymbol{\alpha}_{l,i}\) are provided in Supplementary Material, Sec.~V.

\subsection{From Observation-Vector Errors to Attitude Error}
In the error-free case, the largest eigenvalue of the Davenport gain matrix and its associated unit eigenvector satisfy
\begin{equation}
\mathbf{K}\boldsymbol{q}_1=\lambda_1\boldsymbol{q}_1,\quad \boldsymbol{q}_1^{\mathrm{T}}\boldsymbol{q}_1=1,\quad \lambda_1=\lambda_{\max}(\mathbf{K}).
\label{eq:eigen_nom}
\end{equation}
Let $\delta \mathbf{K}$ be the first-order Davenport gain matrix perturbation caused by observation-vector errors. Linearizing the perturbed eigenvalue equation of the Davenport gain matrix yields the following additive quaternion perturbation, consistent with the analyses in \cite{changErrorAnalysisDavenports2017,ouyangOptimizationBasedStrapdownAttitude2022}:
\begin{equation}
\delta \boldsymbol{q}_1
=
\left(\lambda_1\mathbf{I}_4-\mathbf{K}+\boldsymbol{q}_1\boldsymbol{q}_1^{\mathrm{T}}\right)^{-1}
\left(\mathbf{I}_4-\boldsymbol{q}_1\boldsymbol{q}_1^{\mathrm{T}}\right)\delta \mathbf{K}\boldsymbol{q}_1 .
\label{eq:dq}
\end{equation}
When the largest eigenvalue \(\lambda_1\) is simple, \(\lambda_1\mathbf{I}_4-\mathbf{K}+\boldsymbol{q}_1\boldsymbol{q}_1^{\mathrm{T}}\) is nonsingular. To relate the four-dimensional additive quaternion perturbation to the three-dimensional multiplicative attitude error, a right-multiplicative error quaternion \(\Delta \boldsymbol{q}_1\) is introduced such that \(\boldsymbol{q}_1+\delta \boldsymbol{q}_1=\boldsymbol{q}_1\odot\Delta \boldsymbol{q}_1\), where \(\odot\) denotes quaternion multiplication. Let \(\boldsymbol{q}_1=[s_1,\boldsymbol{\eta}_1^{\mathrm{T}}]^{\mathrm{T}}\). For a small initial-attitude misalignment, \(\Delta \boldsymbol{q}_1\approx[1,\frac{1}{2}(\boldsymbol{\phi}^{i})^{\mathrm{T}}]^{\mathrm{T}}\), yielding the linearized additive-quaternion-to-misalignment mapping
\begin{equation}
\boldsymbol{\phi}^{i}
=2\boldsymbol{\Xi}(\boldsymbol{q}_1)^{\mathrm{T}}\delta \boldsymbol{q}_1,
\quad
\boldsymbol{\Xi}(\boldsymbol{q}_1)^{\mathrm{T}}=
\begin{bmatrix}
-\boldsymbol{\eta}_1 & s_1\mathbf{I}_3-\crossmat{\boldsymbol{\eta}_1}
\end{bmatrix}.
\label{eq:theta_from_q}
\end{equation}
Here, $\boldsymbol{\phi}^{i}$ denotes the Davenport initial-attitude misalignment expressed in the frozen navigation reference frame $n(0)$. Define the corresponding Euler-angle error vector as $\delta\boldsymbol{\theta}=[\delta\theta,\delta\gamma,\delta\psi]^{\mathrm{T}}$. Their first-order relation is $\delta\boldsymbol{\theta}=\mathbf{J}_{\phi}\boldsymbol{\phi}^{i}$, where $\mathbf{J}_{\phi}$ maps the misalignment to the Euler-angle error.

Vectorizing the perturbation matrix $\delta\mathbf{K}$ in \eqref{eq:dq}, substituting it into \eqref{eq:theta_from_q}, and applying the first-order misalignment-to-Euler-angle-error mapping yield the following relationship between the Davenport gain-matrix perturbation and the Euler attitude-error vector:
\begin{equation}
\delta\boldsymbol{\theta}
=\boldsymbol{\Phi}\operatorname{vec}(\delta \mathbf{K}),
\label{eq:phi_map}
\end{equation}
where $\operatorname{vec}(\cdot)$ denotes the column-stacking matrix vectorization operator, $\otimes$ represents the Kronecker product, and the mapping matrix $\boldsymbol{\Phi}$ is given by
\begin{equation}
\begin{aligned}
\boldsymbol{\Phi}
&=
2\mathbf{J}_{\phi}\boldsymbol{\Xi}(\boldsymbol{q}_1)^{\mathrm{T}}
\left(\lambda_1\mathbf{I}_4-\mathbf{K}+\boldsymbol{q}_1\boldsymbol{q}_1^{\mathrm{T}}\right)^{-1}\\
&\quad\times
\left[
\boldsymbol{q}_1^{\mathrm{T}}\otimes \mathbf{I}_4
-\boldsymbol{q}_1^{\mathrm{T}}\otimes(\boldsymbol{q}_1\boldsymbol{q}_1^{\mathrm{T}})
\right].
\end{aligned}
\label{eq:Phi}
\end{equation}
Using the definitions of \(\mathbf{B}\), \(\boldsymbol{z}\), and \(\mathbf{K}\), the observation-vector perturbations induce the following first-order variations:
\begin{equation}
\delta \mathbf{B}
=\sum_{i=1}^{N}w_i
\left(
\delta\boldsymbol{\beta}_i\boldsymbol{\alpha}_i^{\mathrm{T}}
+\boldsymbol{\beta}_i\delta\boldsymbol{\alpha}_i^{\mathrm{T}}
\right),
\label{eq:dB}
\end{equation}
\begin{equation}
\delta\boldsymbol{z}
=\sum_{i=1}^{N}w_i
\left(
\delta\boldsymbol{\beta}_i\times\boldsymbol{\alpha}_i
+\boldsymbol{\beta}_i\times\delta\boldsymbol{\alpha}_i
\right),
\label{eq:dZ}
\end{equation}
\begin{equation}
\delta \mathbf{K}=
\begin{bmatrix}
\operatorname{tr}(\delta \mathbf{B}) & \delta\boldsymbol{z}^{\mathrm{T}}\\
\delta\boldsymbol{z} & \delta \mathbf{B}+\delta \mathbf{B}^{\mathrm{T}}-\operatorname{tr}(\delta \mathbf{B})\mathbf{I}_3
\end{bmatrix}.
\label{eq:dK}
\end{equation}
Equations \eqref{eq:dB}--\eqref{eq:dK} establish a first-order linear mapping from the observation-vector perturbations to \(\delta \mathbf{K}\). Therefore, by vectorizing the matrix equation in \eqref{eq:dK}, the linear coefficients associated with $\delta\boldsymbol{\beta}_i$ and $\delta\boldsymbol{\alpha}_i$ can be explicitly extracted, and the corresponding Jacobian matrices are defined as
\begin{equation}
\operatorname{vec}(\delta \mathbf{K})
=
\sum_{i=1}^{N}
\left(
\mathbf{J}_{\beta,i}\delta\boldsymbol{\beta}_i
+\mathbf{J}_{\alpha,i}\delta\boldsymbol{\alpha}_i
\right)
=\mathbf{J}_{\mathrm{K}}\delta\boldsymbol{y},
\label{eq:JK}
\end{equation}
where $\delta\boldsymbol{y}=[\delta\boldsymbol{\beta}_1^{\mathrm{T}},\delta\boldsymbol{\alpha}_1^{\mathrm{T}},\ldots,\delta\boldsymbol{\beta}_N^{\mathrm{T}},\delta\boldsymbol{\alpha}_N^{\mathrm{T}}]^{\mathrm{T}}$, $\mathbf{J}_{\beta,i}$ and $\mathbf{J}_{\alpha,i}$ are the Jacobian matrices of $\operatorname{vec}(\mathbf{K})$ with respect to the $i$th observation pair, and
\begin{equation}
\mathbf{J}_{\mathrm{K}}
=
\begin{bmatrix}
\mathbf{J}_{\beta,1} & \mathbf{J}_{\alpha,1} & \cdots & \mathbf{J}_{\beta,N} & \mathbf{J}_{\alpha,N}
\end{bmatrix}
\in\mathbb{R}^{16\times6N}.
\label{eq:JK_def}
\end{equation}
Combining \eqref{eq:phi_map} and \eqref{eq:JK} yields
\begin{equation}
\delta\boldsymbol{\theta}
=\boldsymbol{\Phi}\mathbf{J}_{\mathrm{K}}\delta\boldsymbol{y}.
\label{eq:theta_y}
\end{equation}
The derivations of \eqref{eq:dq} and \eqref{eq:theta_from_q}, the relationship between \(\Delta\boldsymbol{q}_1\) and \(\boldsymbol{\phi}^{i}\), and the constructions of \(\mathbf{J}_{\phi}\), \(\mathbf{J}_{\beta,i}\), and \(\mathbf{J}_{\alpha,i}\) are provided in Supplementary Material, Sec.~VI.

\subsection{Deterministic and Stochastic Error Components}
The input errors are decomposed into a deterministic component $\mathcal{S}$ and a stochastic component $\mathcal{R}$:
\begin{equation}
\begin{aligned}
\mathcal{S}:\;&
\delta\boldsymbol{\omega}_{ib}^b=\boldsymbol{\epsilon}^b,\quad
\delta\boldsymbol{f}^b=\boldsymbol{\nabla}^b,\quad
\boldsymbol{l}^b\neq\boldsymbol{0},\\
&
\delta\boldsymbol{v}_{GNSS}^n=\boldsymbol{0},\\
\mathcal{R}:\;&
\delta\boldsymbol{\omega}_{ib}^b=\boldsymbol{w}_g,\quad
\delta\boldsymbol{f}^b=\boldsymbol{w}_a,\quad
\boldsymbol{l}^b=\boldsymbol{0},\\
&
\delta\boldsymbol{v}_{GNSS}^n=\boldsymbol{w}_v .
\end{aligned}
\label{eq:tracks}
\end{equation}
Any slowly varying or systematic input errors not captured by the white-noise terms in \eqref{eq:meas_model} should be included in \(\mathcal{S}\) or represented by an extended stochastic model.

The deterministic errors are treated as known constants. Their combined contributions across all sliding windows yield the deterministic attitude offset
\begin{equation}
\delta\boldsymbol{\theta}_s
=
\boldsymbol{\Phi}
\sum_{i=1}^{N}
\mathbf{J}_{\alpha,i}
\left[
\delta\boldsymbol{\alpha}_{g,i}(\boldsymbol{\epsilon}^b)
+\delta\boldsymbol{\alpha}_{a,i}(\boldsymbol{\nabla}^b)
-\delta\boldsymbol{\alpha}_{l,i}(\boldsymbol{l}^b)
\right].
\label{eq:theta_system}
\end{equation}

For the stochastic component, gyroscope-noise-induced accumulated attitude projection errors grow over time. Adjacent windows can also share endpoint GNSS velocity samples and overlapping IMU integration intervals. Consequently, observation-vector errors in different windows are temporally correlated, and the complete observation covariance $\mathbf{P}_{y,r}$ is generally a high-dimensional dense matrix. To avoid explicitly constructing the dense covariance matrix of the stacked window errors, the order of summation is interchanged, thereby reorganizing the contributions by global epoch. For the discrete noise samples indexed by \(k=1,\ldots,K_{\max}\), the stochastic attitude error can be reorganized as
\begin{equation}
\begin{aligned}
\delta\boldsymbol{\theta}_r
&=
\boldsymbol{\Phi}
\sum_{i=1}^{N}
\left(
\mathbf{J}_{\beta,i}\delta\boldsymbol{\beta}_{v,i}
+\mathbf{J}_{\alpha,i}\delta\boldsymbol{\alpha}_{a,i}
+\mathbf{J}_{\alpha,i}\delta\boldsymbol{\alpha}_{g,i}
\right)\\
&=
\boldsymbol{\Phi}
\sum_{k=1}^{K_{\max}}
\left(
\mathbf{X}_k\boldsymbol{w}_{a,k}
+\mathbf{Y}_k\boldsymbol{w}_{g,k}
+\mathbf{Z}_k\boldsymbol{w}_{v,k}
\right).
\end{aligned}
\label{eq:theta_random_epoch}
\end{equation}
Here, \(\boldsymbol{w}_{a,k}\), \(\boldsymbol{w}_{g,k}\), and \(\boldsymbol{w}_{v,k}\) are the accelerometer, gyroscope, and GNSS velocity white-noise samples at epoch \(k\), respectively. The matrices \(\mathbf{X}_k\), \(\mathbf{Y}_k\), and \(\mathbf{Z}_k\) are the global propagation coefficients from the corresponding noise sources to \(\operatorname{vec}(\mathbf{K})\), whose explicit constructions are provided in Supplementary Material, Sec.~VII.

Since the white-noise samples at different discrete epochs are mutually independent, cross-epoch terms vanish in \(\mathbf{P}_{\theta,r}=E[\delta\boldsymbol{\theta}_r\delta\boldsymbol{\theta}_r^{\mathrm{T}}]\). The attitude-error covariance is obtained by summing the single-epoch covariance contributions:
\begin{equation}
\begin{aligned}
\mathbf{P}_{\theta,r}
&=
\boldsymbol{\Phi}
\left[
\sum_{k=1}^{K_{\max}}
\left(
\mathbf{X}_k\mathbf{Q}_{a,k}\mathbf{X}_k^{\mathrm{T}}
+\mathbf{Y}_k\mathbf{Q}_{g,k}\mathbf{Y}_k^{\mathrm{T}}
\right.\right.\\
&\qquad\left.\left.
+\mathbf{Z}_k\mathbf{R}_{v,k}\mathbf{Z}_k^{\mathrm{T}}
\right)
\right]
\boldsymbol{\Phi}^{\mathrm{T}} .
\end{aligned}
\label{eq:theta_cov}
\end{equation}
Here, \(\mathbf{Q}_{a,k}\in\mathbb{R}^{6\times6}\) is the covariance of the two stacked half-period accelerometer velocity-increment noise samples, whereas \(\mathbf{Q}_{g,k}\) and \(\mathbf{R}_{v,k}\) are the gyroscope and aiding-velocity noise covariance matrices, respectively. For uncorrelated, equal-variance GNSS velocity noise, \(\mathbf{R}_{v,k}=\sigma_v^2\mathbf{I}_3\), where \(\sigma_v\) is the standard deviation. The attitude error evaluation of sliding-window OBA is then
\begin{equation}
E[\delta\boldsymbol{\theta}]
=\delta\boldsymbol{\theta}_s,
\quad
\operatorname{Cov}(\delta\boldsymbol{\theta})
=\mathbf{P}_{\theta,r}.
\label{eq:final_eval}
\end{equation}
Equation \eqref{eq:final_eval} indicates that systematic errors dictate the deterministic attitude offset, while stochastic noise determines the statistical envelope around it. The $3\sigma$ bounds in pitch, roll, and heading can be obtained from $\delta\boldsymbol{\theta}_s\pm3\sqrt{\operatorname{diag}(\mathbf{P}_{\theta,r})}$.

Table~\ref{tab:error_algorithm} summarizes the computational procedure; detailed derivations of \(\delta\boldsymbol{\theta}_s\) and \(\mathbf{P}_{\theta,r}\) are provided in Supplementary Material, Sec.~VII.

\begin{table}[!t]
\caption{Procedure for Sliding-Window OBA Error Evaluation}
\label{tab:error_algorithm}
\centering
\footnotesize
\begin{tabular}{@{}>{\raggedright\arraybackslash}p{0.16\columnwidth}>{\raggedright\arraybackslash}p{0.76\columnwidth}@{}}
\hline
Step & Key operations and equations \\
\hline
Step 1 & Set $\boldsymbol{\epsilon}^b$, $\boldsymbol{\nabla}^b$, $\boldsymbol{l}^b$, $\mathbf{Q}_{a,k}$, $\mathbf{Q}_{g,k}$, $\mathbf{R}_{v,k}$, and window indices $(m_i,M_i)$. \\
Step 2 & Construct $(\boldsymbol{\alpha}_i,\boldsymbol{\beta}_i)$ via \eqref{eq:alpha_disc}, \eqref{eq:beta_disc}, and form $\mathbf{B}$, $\boldsymbol{z}$, and $\mathbf{K}$ via \eqref{eq:BZ}, \eqref{eq:K}. \\
Step 3 & Solve $(\lambda_1,\boldsymbol{q}_1)$ via \eqref{eq:qmax}, compute $\boldsymbol{\Phi}$ via \eqref{eq:Phi}, and assemble $\mathbf{J}_{\mathrm{K}}$ via \eqref{eq:JK}, \eqref{eq:JK_def}. \\
Step 4 & Compute $\delta\boldsymbol{\alpha}_{a,i}(\boldsymbol{\nabla}^b)$, $\delta\boldsymbol{\alpha}_{g,i}(\boldsymbol{\epsilon}^b)$, and $\delta\boldsymbol{\alpha}_{l,i}(\boldsymbol{l}^b)$ by \eqref{eq:alpha_accel}, \eqref{eq:alpha_gyro}, and \eqref{eq:alpha_lever}; evaluate $\delta\boldsymbol{\theta}_s$ by \eqref{eq:theta_system}. \\
Step 5 & Evaluate \(\mathbf{P}_{\theta,r}\) by \eqref{eq:theta_cov}; obtain the mean, covariance, and \(\pm3\sigma\) bounds by \eqref{eq:final_eval}. \\
\hline
\end{tabular}
\end{table}

\section{Simulation Validation}
\begin{figure}[tbp]
\centering
\includegraphics[width=0.98\columnwidth]{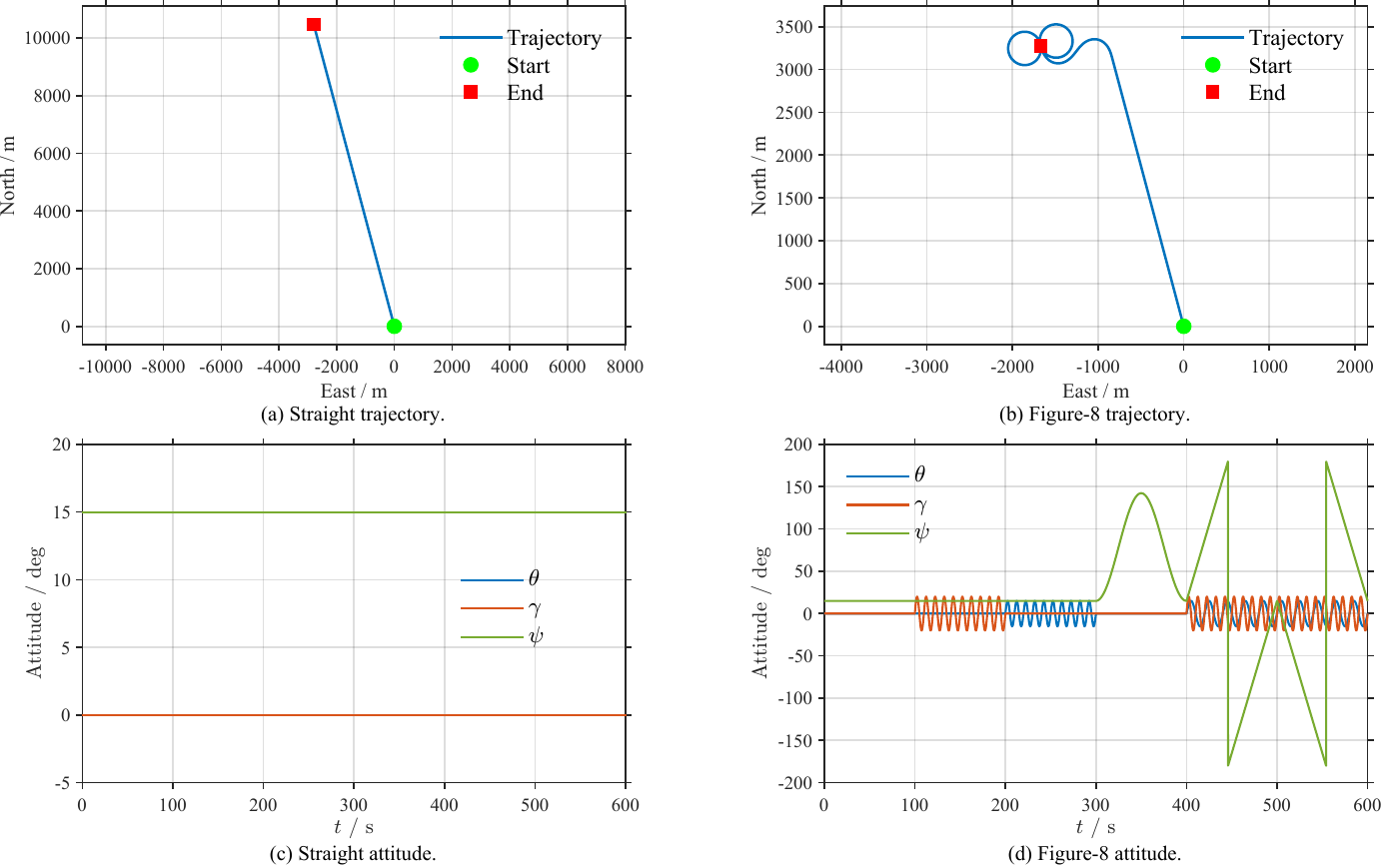}
\caption{Straight and Figure-8 trajectories and attitude profiles used to represent weakly and strongly excited motion conditions.}
\label{fig:trajectory_truth}
\end{figure}

To validate the deterministic error offset and attitude error covariance models, a sliding-window OBA simulation is conducted under two contrasting trajectories. As shown in Fig.~\ref{fig:trajectory_truth}, the straight trajectory represents weakly excited motion, where slowly varying specific-force and velocity directions make heading weakly observable. The Figure-8 trajectory represents strongly excited multi-axis maneuvering, where acceleration, deceleration, and turning diversify the observation-vector geometry. Table~\ref{tab:settings} lists the main simulation settings, with the inertial-sensor random-walk coefficients converted to equivalent discrete white-noise samples at \(f_s=100~\mathrm{Hz}\). The simulation is organized into three parts: deterministic error injection, stochastic Monte Carlo covariance evaluation, and simultaneous combined-error injection. Each Monte Carlo evaluation uses 500 trials.

\begin{table}[!t]
\caption{Simulation Settings and Injected Error Parameters}
\label{tab:settings}
\centering
\begin{tabular}{@{}ll@{}}
\toprule
Parameter & Value \\
\midrule
Sampling rate $f_s$ & $100~\mathrm{Hz}$ \\
Simulation duration $T_{\mathrm{sim}}$ & $600~\mathrm{s}$ \\
Window length $T_w$ & $10~\mathrm{s}$ \\
Monte Carlo trials $N_{\mathrm{MC}}$ & $500$ \\
Gyroscope bias $\boldsymbol{\epsilon}^b$ & $[0.5,-0.4,0.3]~\mathrm{deg/h}$ \\
Accelerometer bias $\boldsymbol{\nabla}^b$ & $[300,-200,250]~\mu\mathrm{g}$ \\
Lever arm $\boldsymbol{l}^b$ & $[0.8,-0.5,0.3]~\mathrm{m}$ \\
GNSS velocity-noise standard deviation $\sigma_v$ & $0.1~\mathrm{m/s}$ \\
Gyroscope angle random walk (ARW) & $0.06~\mathrm{deg}/\sqrt{\mathrm{h}}$ \\
Accelerometer velocity random walk (VRW) & $1.2\times10^{-2}~\mathrm{m/s}/\sqrt{\mathrm{h}}$ \\
\bottomrule
\end{tabular}
\end{table}

\subsection{Deterministic Error}
\begin{figure*}[!t]
\centering
\includegraphics[width=0.80\textwidth]{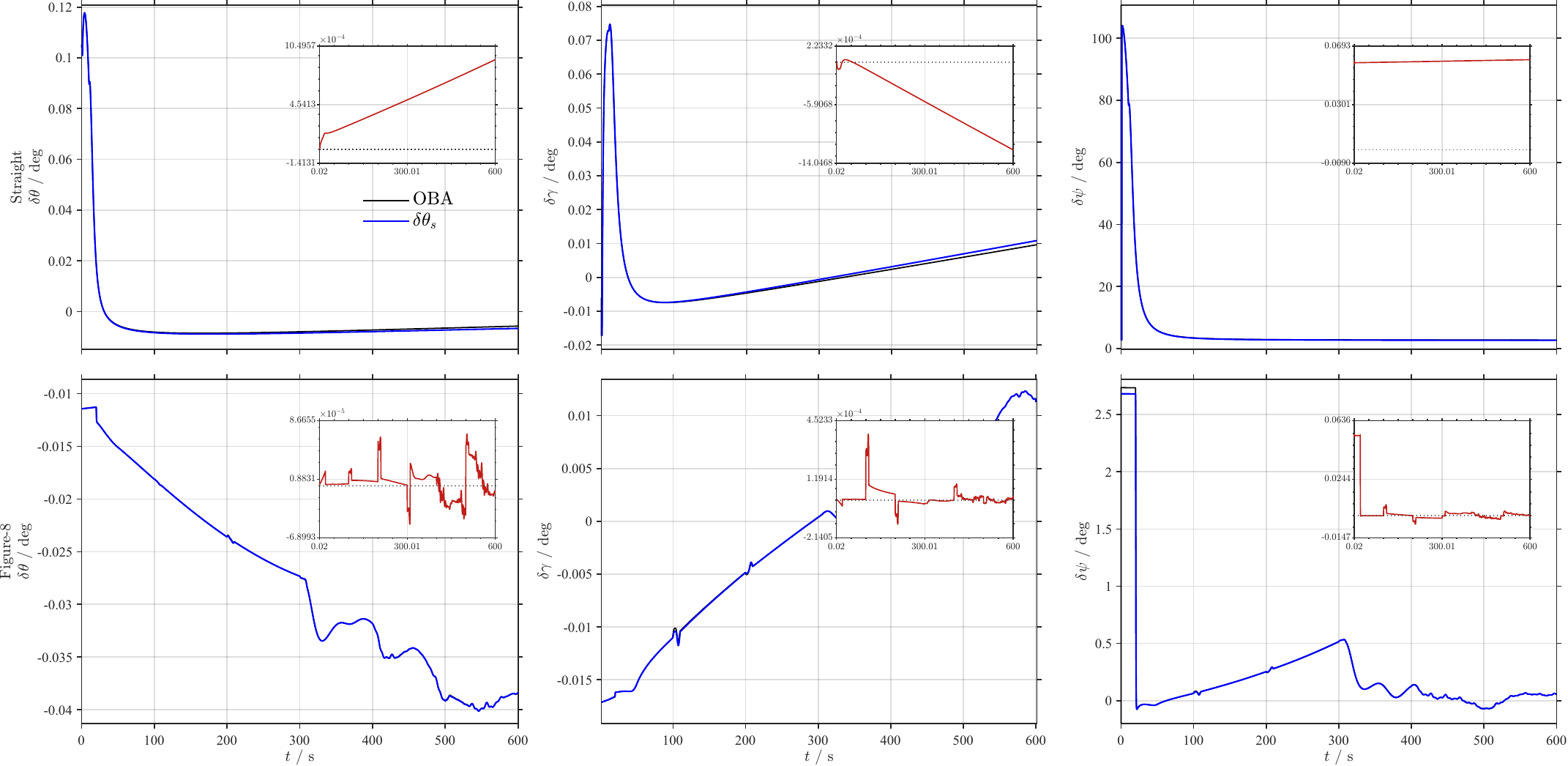}
\caption{Deterministic OBA attitude errors and first-order analytical predictions under simultaneous deterministic-error injection. Insets show instantaneous prediction residuals.}
\label{fig:initial_error}
\end{figure*}

The deterministic experiment compares OBA initial-attitude errors under deterministic-error injection with the first-order prediction in \eqref{eq:theta_system}. The attitude-error prediction residual is
\begin{equation}
\boldsymbol{r}_k
=
\delta\boldsymbol{\theta}_{\mathrm{OBA}}(t_k)
-\delta\boldsymbol{\theta}_s(t_k),
\qquad k=1,\ldots,K_{\max}.
\label{eq:det_residual}
\end{equation}
The final residual vector, the componentwise RMSE vector, and their corresponding three-axis norms are defined as
\begin{equation}
\begin{aligned}
\boldsymbol{r}_{K_{\max}}
&=
\delta\boldsymbol{\theta}_{\rm OBA}(t_{K_{\max}})
-\delta\boldsymbol{\theta}_s(t_{K_{\max}}),\\
\boldsymbol{r}_{\rm RMSE}
&=
\sqrt{\frac{1}{K_{\max}}
\sum_{k=1}^{K_{\max}}
\operatorname{diag}\!\left(
\boldsymbol{r}_k\boldsymbol{r}_k^{\mathrm{T}}
\right)},\\
e_{\rm final}
&=\left\|\boldsymbol{r}_{K_{\max}}\right\|_2,\\
e_{\rm RMSE}
&=\sqrt{\frac{1}{K_{\max}}
\sum_{k=1}^{K_{\max}}
\left\|\boldsymbol{r}_k\right\|_2^2}.
\end{aligned}
\label{eq:det_residual_metrics}
\end{equation}
The square root in the definition of \(\boldsymbol{r}_{\rm RMSE}\) is taken elementwise, and \(e_{\rm RMSE}=\|\boldsymbol{r}_{\rm RMSE}\|_2\). Table~\ref{tab:system} reports \(e_{\rm final}\) and \(e_{\rm RMSE}\) for each trajectory and deterministic-error case, whereas Fig.~\ref{fig:initial_error} shows the attitude errors and instantaneous residuals under simultaneous deterministic-error injection.

\begin{table}[t]
\caption{Deterministic Initial-Attitude Prediction Residual Norms}
\label{tab:system}
\centering
\footnotesize
\setlength{\tabcolsep}{1.8pt}
\begin{tabular}{llcc}
\toprule
Trajectory & Source & \(e_{\rm final}\) / deg & \(e_{\rm RMSE}\) / deg \\
\midrule
Straight & $\boldsymbol{\epsilon}^b$ & $6.132\times10^{-2}$ & $6.026\times10^{-2}$ \\
Straight & $\boldsymbol{\nabla}^b$ & $6.508\times10^{-6}$ & $4.338\times10^{-5}$ \\
Straight & $\boldsymbol{l}^b$ & $2.289\times10^{-7}$ & $2.477\times10^{-5}$ \\
Straight & $\boldsymbol{\epsilon}^b+\boldsymbol{\nabla}^b+\boldsymbol{l}^b$ & $6.029\times10^{-2}$ & $5.925\times10^{-2}$ \\
Figure-8 & $\boldsymbol{\epsilon}^b$ & $6.862\times10^{-5}$ & $1.001\times10^{-2}$ \\
Figure-8 & $\boldsymbol{\nabla}^b$ & $8.161\times10^{-6}$ & $8.650\times10^{-5}$ \\
Figure-8 & $\boldsymbol{l}^b$ & $3.815\times10^{-4}$ & $1.658\times10^{-3}$ \\
Figure-8 & $\boldsymbol{\epsilon}^b+\boldsymbol{\nabla}^b+\boldsymbol{l}^b$ & $4.350\times10^{-4}$ & $9.940\times10^{-3}$ \\
\bottomrule
\end{tabular}
\end{table}

The nearly overlapping analytical and numerical curves in Fig.~\ref{fig:initial_error} indicate that the proposed model accurately predicts the dominant deterministic offsets induced by sensor biases and lever-arm effects. Under the straight trajectory, weak heading observability causes the vertical-axis gyroscope bias to dominate the deterministic heading error, resulting in a large heading offset under simultaneous deterministic-error injection. In contrast, the Figure-8 maneuver improves the observation geometry and suppresses this deterministic heading offset. Quantitatively, simultaneous deterministic-error injection under the straight trajectory produces a heading offset exceeding 2.5 deg, whereas the Figure-8 maneuver suppresses it below 0.1 deg. Table~\ref{tab:system} further shows that the dominant contributor to the final residual norm shifts from gyroscope bias under the straight trajectory to the lever-arm effect under the Figure-8 maneuver, owing to the increased excitation of the lever-arm effect by angular motion. The residuals shown in the inset of Fig.~\ref{fig:initial_error} and the quantitative results in Table~\ref{tab:system} further show that the prediction residuals are much smaller than the corresponding attitude offsets, confirming that the first-order model captures the deterministic error evolution.

\subsection{Stochastic Error}
To validate the stochastic error propagation, gyroscope, accelerometer, and GNSS velocity noise are injected separately and jointly. The standard-deviation ratio \(\widehat{\sigma}_{\rm MC}/\sigma_{\rm pred}\) compares the sample standard deviation from 500 Monte Carlo trials with that predicted by \eqref{eq:theta_cov}; a value close to unity indicates agreement between the Monte Carlo dispersion and the analytical covariance.

\begin{figure}[!t]
\centering
\begin{minipage}[t]{0.48\columnwidth}
\centering
\includegraphics[width=0.86\linewidth]{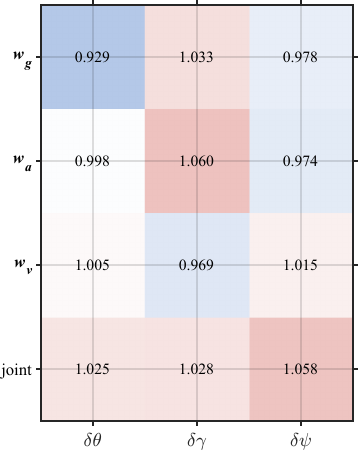}
\par\vspace{0.1ex}
\footnotesize (a) Straight trajectory.
\end{minipage}\hfill
\begin{minipage}[t]{0.48\columnwidth}
\centering
\includegraphics[width=\linewidth]{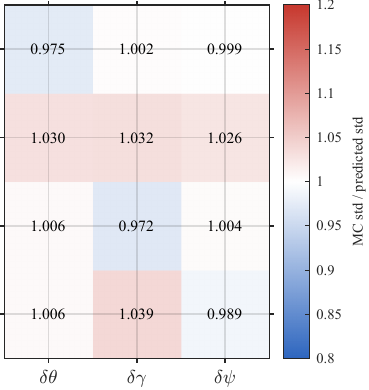}
\par\vspace{0.1ex}
\footnotesize (b) Figure-8 trajectory.
\end{minipage}
\caption{Standard-deviation ratio heatmaps \(\widehat{\sigma}_{\rm MC}/\sigma_{\rm pred}\) under stochastic-noise injection for the straight and Figure-8 trajectories.}
\label{fig:covariance_heatmap}
\end{figure}

As shown in Fig.~\ref{fig:covariance_heatmap}, the analytical covariance model maintains all three-axis standard-deviation ratios between 0.929 and 1.060 across both trajectories. Weak heading observability under the straight trajectory amplifies the heading uncertainty, whereas the Figure-8 maneuver improves the observation geometry and reduces it; in both cases, the standard-deviation ratios remain close to unity. These results verify that the proposed covariance model accurately evaluates stochastic attitude-error statistics under different motion conditions.

\begin{figure*}[!b]
\centering
\includegraphics[width=0.80\textwidth]{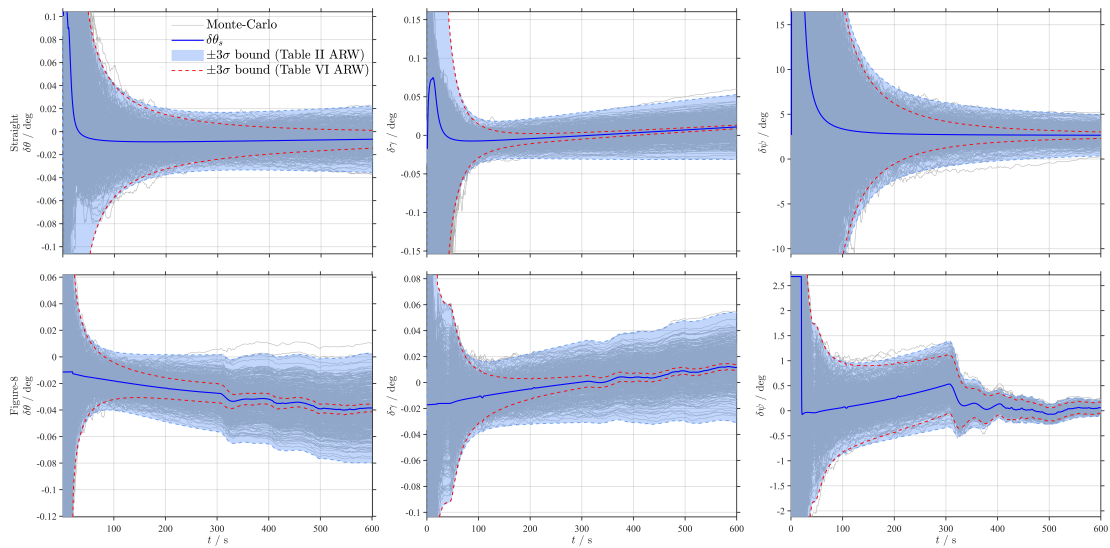}
\caption{Monte Carlo attitude errors and analytical \(\pm3\sigma\) envelopes under simultaneous combined-error injection. Blue and red dashed lines denote bounds computed with the ARW settings in Table~\ref{tab:settings} and Table~\ref{tab:field_inputs}, respectively.}
\label{fig:mc_envelope}
\end{figure*}

\subsection{Combined-Error Envelope}
The final simulation injects all deterministic and stochastic error sources. The deterministic model provides the time-varying offset \(\delta\boldsymbol{\theta}_s\), while the covariance model generates the dynamic \(\pm3\sigma\) envelope. The Monte Carlo sample-mean attitude-error vector is
\begin{equation}
\overline{\delta\boldsymbol{\theta}}_{\rm MC,OBA}(t_k)
=
\frac{1}{N_{\rm MC}}
\sum_{\ell=1}^{N_{\rm MC}}
\delta\boldsymbol{\theta}_{\rm OBA}^{(\ell)}(t_k),
\label{eq:mc_mean_attitude}
\end{equation}
where \(\ell\) denotes the Monte Carlo trial index. Using \eqref{eq:mc_mean_attitude}, the final Monte Carlo mean-residual vector is defined as
\begin{equation}
\overline{\boldsymbol{r}}_{K_{\max}}
=
\overline{\delta\boldsymbol{\theta}}_{\rm MC,OBA}(t_{K_{\max}})
-\delta\boldsymbol{\theta}_s(t_{K_{\max}}),
\label{eq:mean_residual}
\end{equation}
Table~\ref{tab:random} reports the pitch, roll, and heading components of \(\overline{\boldsymbol{r}}_{K_{\max}}\). The final-epoch coverage is the fraction of 500 Monte Carlo trials lying within the analytical \(\pm3\sigma\) bounds.

\begin{table}[t]
\caption{Statistical Summary Under Simultaneous Combined-Error Injection}
\label{tab:random}
\centering
\footnotesize
\setlength{\tabcolsep}{1.6pt}
\begin{tabular}{llccc}
\toprule
Trajectory & Axis &
\begin{tabular}[c]{@{}c@{}}Component of\\ \(\overline{\boldsymbol{r}}_{K_{\max}}\) / deg\end{tabular} &
\begin{tabular}[c]{@{}c@{}}Std. ratio\\ \(\widehat{\sigma}_{\mathrm{MC}}/\sigma_{\mathrm{pred}}\)\end{tabular} &
\begin{tabular}[c]{@{}c@{}}Final-epoch\\coverage\end{tabular} \\
\midrule
Straight & Pitch & $9.00\times10^{-4}$ & $1.036$ & $0.994$ \\
Straight & Roll & $-5.81\times10^{-5}$ & $0.944$ & $0.998$ \\
Straight & Heading & $6.75\times10^{-2}$ & $0.998$ & $0.996$ \\
Figure-8 & Pitch & $-1.94\times10^{-4}$ & $1.018$ & $0.998$ \\
Figure-8 & Roll & $1.27\times10^{-3}$ & $0.942$ & $0.998$ \\
Figure-8 & Heading & $9.23\times10^{-4}$ & $1.017$ & $0.996$ \\
\bottomrule
\end{tabular}
\end{table}

As shown in Fig.~\ref{fig:mc_envelope} and Table~\ref{tab:random}, the analytical envelopes effectively bound the 500 Monte Carlo trajectories. The empirical coverage ratios remain above \(0.994\), and the standard-deviation ratios remain within 0.942--1.036, indicating that the predicted bounds are both reliable and tight. The components of \(\overline{\boldsymbol{r}}_{K_{\max}}\) are on the order of \(10^{-5}\) to \(10^{-2}\) deg, confirming that the deterministic offsets remain accurate under simultaneous combined-error injection. The Figure-8 turning maneuver near \(t=300\) s further improves the heading observation geometry and accelerates heading-error convergence. Overall, these results confirm that the proposed first-order model accurately predicts both the deterministic offset and the associated attitude-error covariance under combined deterministic and stochastic error injection.

The slight late-stage expansion of the horizontal covariance envelopes in Fig.~\ref{fig:mc_envelope} is caused by the gyroscope ARW and its coupling with the gravity-dominated specific force through \(\mathbf{Y}_k\), whose explicit construction is provided in Supplementary Material, Sec.~VII. Because gravity dominates \(\boldsymbol{f}^b\), the cross-product term in \(\mathbf{Y}_k\) amplifies the horizontal components of \(\delta\boldsymbol{\alpha}_g\) more strongly than the vertical component, explaining why the expansion mainly appears in pitch and roll. When the ARW is reduced from the value used in Table~\ref{tab:settings} to the NS260 level in Table~\ref{tab:field_inputs}, the critical point of divergence is delayed, and the horizontal error envelopes remain monotonically convergent within the 600 s duration. This confirms that the envelope expansion is primarily attributable to the sensor-noise magnitude rather than to a breakdown of the propagation model.

\section{Vehicle Field-Test Validation}
\begin{figure}[!t]
\centering
\setlength{\tabcolsep}{0pt}
\begin{tabular}{@{}c@{\hspace{0.01\columnwidth}}c@{}}
\begin{minipage}[t]{0.495\linewidth}
\centering
\includegraphics[width=\linewidth,trim=10 10 10 10,clip]{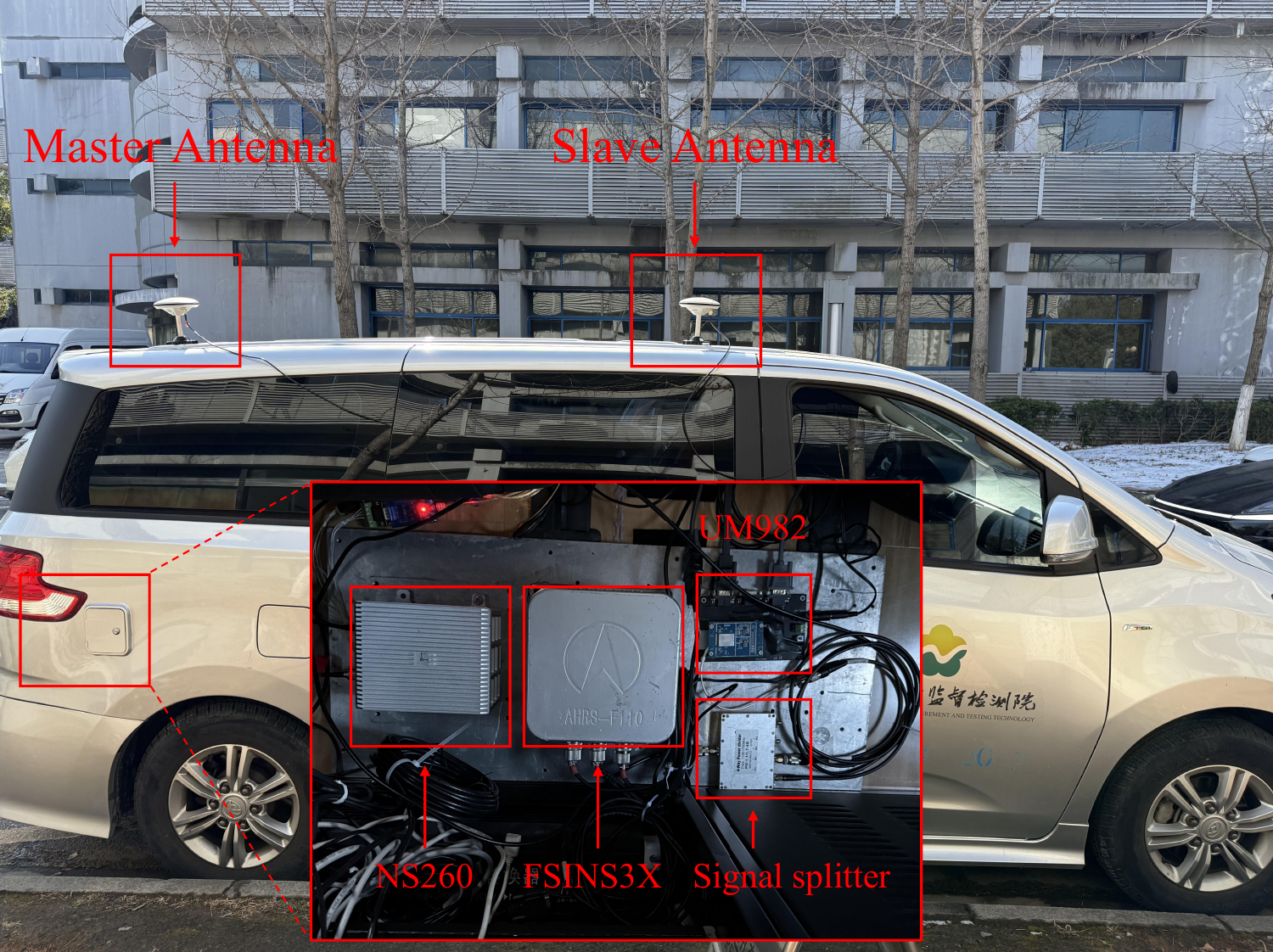}
\vspace{0.1ex}
\makebox[\linewidth][c]{\footnotesize (a) Vehicle platform.}
\end{minipage} &
\begin{minipage}[t]{0.495\linewidth}
\centering
\includegraphics[width=\linewidth,trim=12 12 12 12,clip]{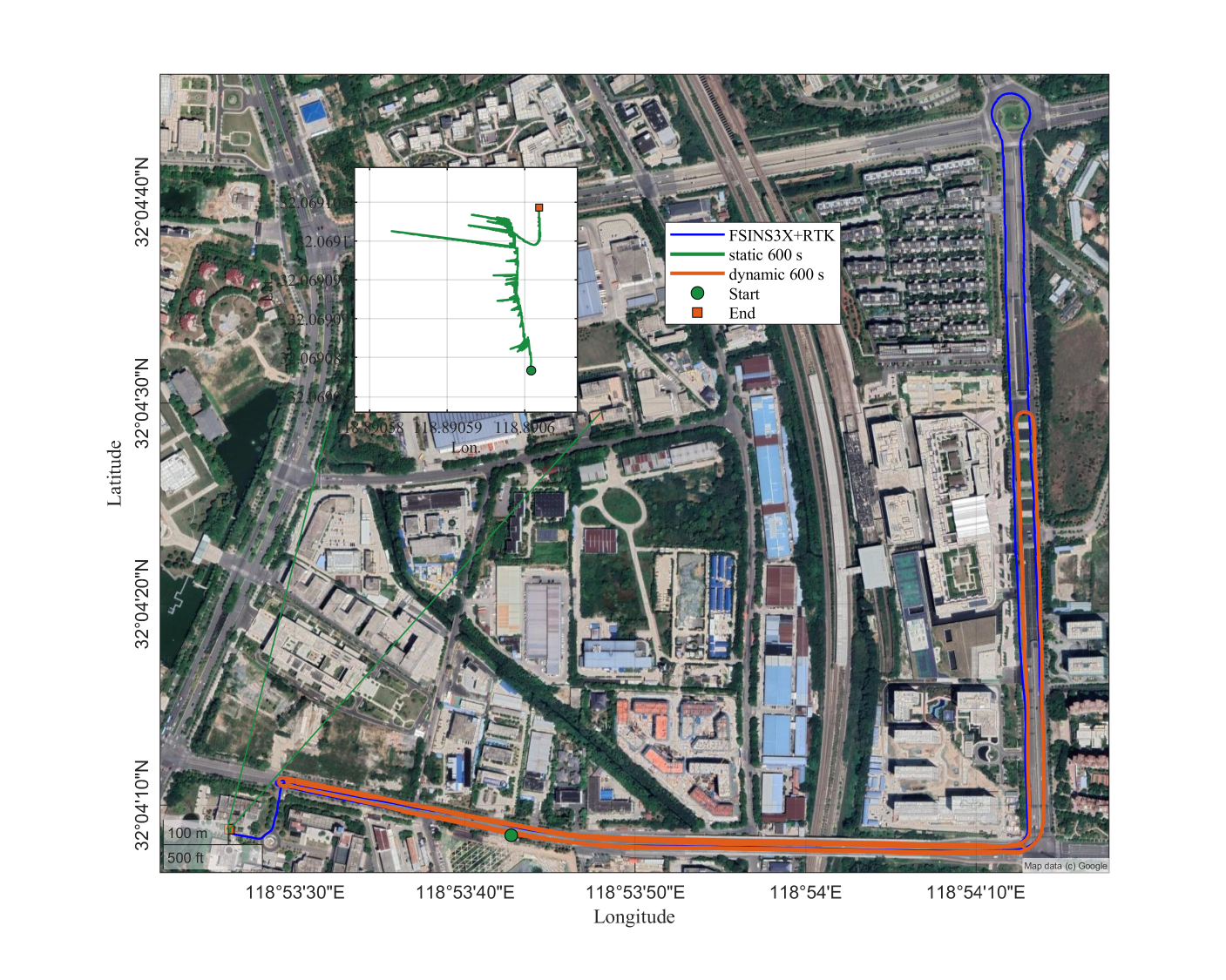}
\vspace{0.1ex}
\makebox[\linewidth][c]{\footnotesize (b) Full-route trajectory.}
\end{minipage}
\\[0.2em]

\begin{minipage}[t]{0.495\linewidth}
\centering
\includegraphics[width=\linewidth,trim=4 4 4 4,clip]{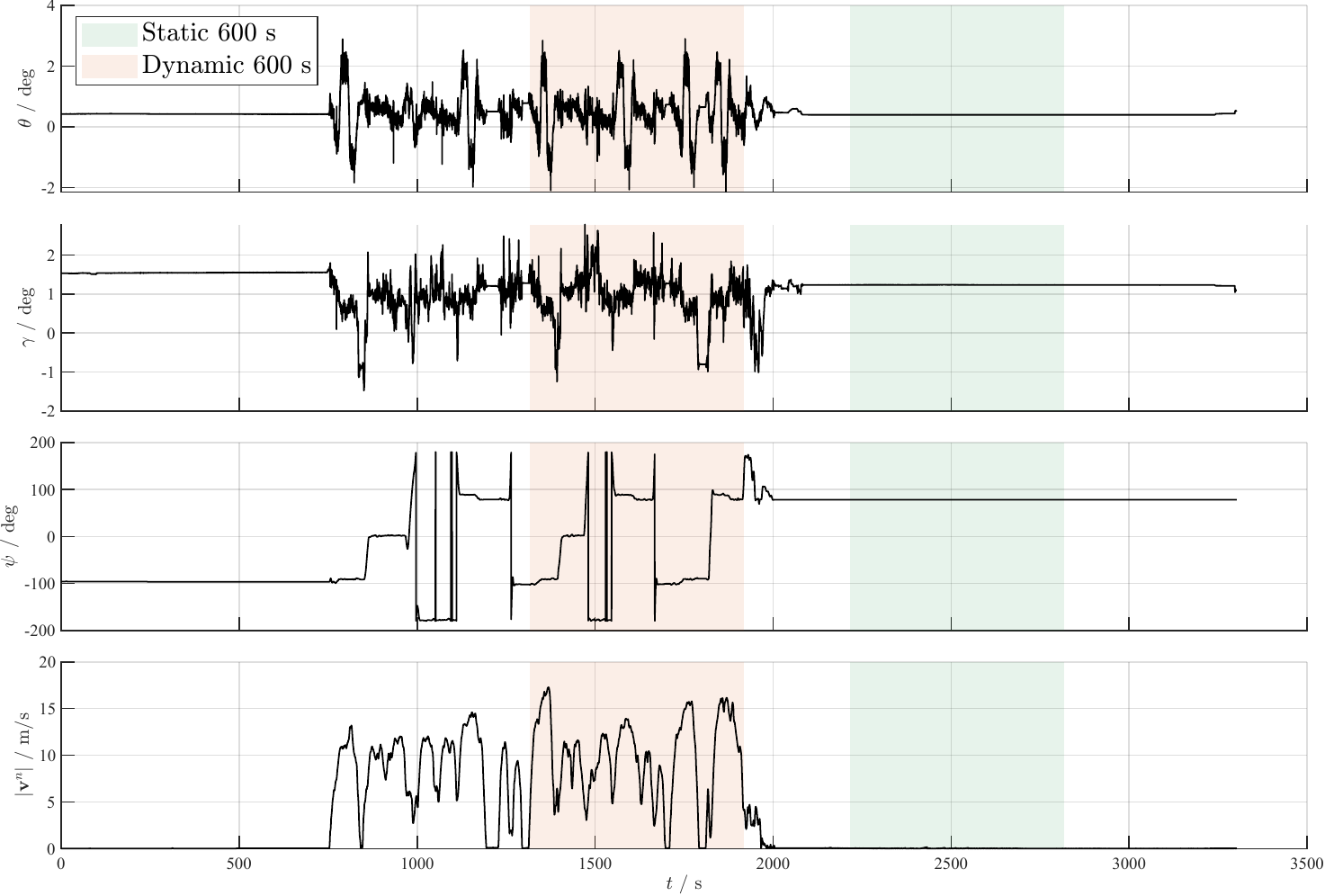}
\vspace{0.1ex}
\makebox[\linewidth][c]{\footnotesize (c) Reference motion profiles.}
\end{minipage} &
\begin{minipage}[t]{0.495\linewidth}
\centering
\includegraphics[width=\linewidth,trim=4 4 4 4,clip]{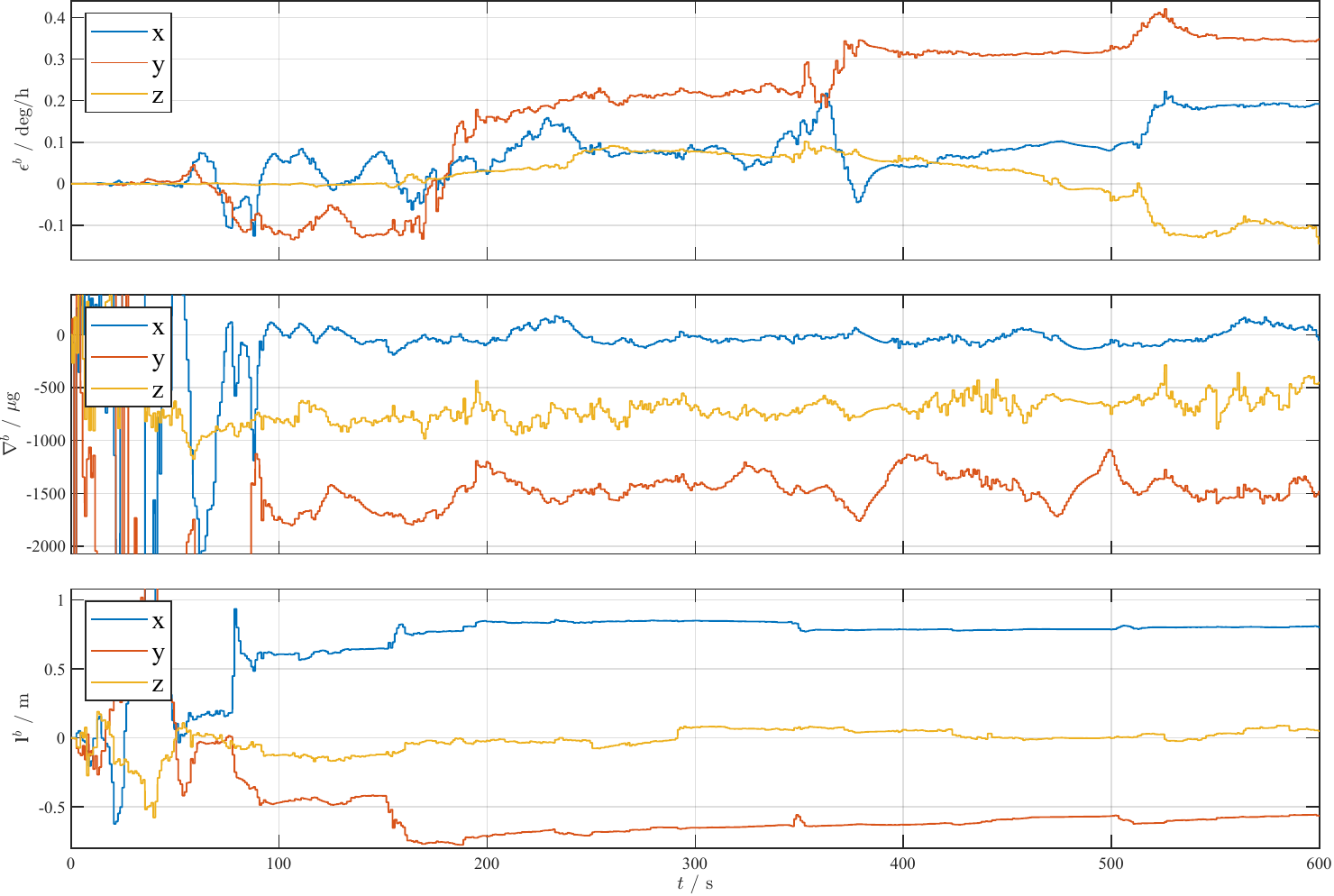}
\vspace{0.1ex}
\makebox[\linewidth][c]{\footnotesize (d) Filter convergence.}
\end{minipage}
\end{tabular}
\caption{Field-test setup, full-route trajectory, reference motion profiles, and deterministic-error identification.}
\label{fig:field_overview}
\end{figure}

To evaluate whether the proposed error-propagation model remains valid with real sensor data, a vehicle field test was conducted using two rigidly mounted navigation systems, as shown in Fig.~\ref{fig:field_overview}(a). The FSINS3X/real-time kinematic (RTK) system provides the reference attitude, velocity, and position baseline, while the NS260 with single-point GNSS provides the raw inertial measurements and GNSS velocity for sliding-window OBA. Table~\ref{tab:field_specs} summarizes the hardware configuration.

Approximately 3200 s of FSINS3X/RTK reference solutions and NS260/GNSS raw data were recorded, as shown in Fig.~\ref{fig:field_overview}(b) and (c). A 600 s dynamic segment and a 600 s static segment were extracted. The dynamic segment contains strong velocity and attitude variations, providing sufficient observability for the deterministic error sources, whereas the static segment represents poor excitation and is used to examine whether the covariance model correctly predicts the growth of the heading-error envelope.

\begin{table}[!t]
\caption{Hardware Configuration and System Specifications}
\label{tab:field_specs}
\centering
\footnotesize
\setlength{\tabcolsep}{2.5pt}
\renewcommand{\arraystretch}{1.25}
\begin{tabular}{@{}>{\raggedright\arraybackslash}p{0.25\columnwidth}>{\raggedright\arraybackslash}p{0.28\columnwidth}>{\raggedright\arraybackslash}p{0.38\columnwidth}@{}}
\toprule
System role & Configuration & Integrated accuracy \\
\midrule
\textbf{Reference baseline} &
FSINS3X + RTK\newline
($200~\mathrm{Hz}$) &
Att: $0.02^\circ/0.02^\circ/0.06^\circ$\newline
Vel: $0.03~\mathrm{m/s}$\newline
Pos: $5~\mathrm{cm}$ (H), $10~\mathrm{cm}$ (V) \\
\textbf{Test system} &
NS260 + GNSS\newline
($200~\mathrm{Hz}$) &
Error parameters\newline
detailed in Table~\ref{tab:field_inputs}. \\
\bottomrule
\end{tabular}
\end{table}

The purpose of the field test is to examine whether the proposed analytical error-propagation model can predict the actual OBA initial-attitude errors. Therefore, the analytical model requires input parameters consistent with the actual sensor characteristics. These inputs are extracted from the raw experimental data, rather than taken directly from nominal manufacturer specifications or coarse manual estimates, so as to reduce prediction residuals caused by input-parameter mismatch. For practical online deployment, enlarged empirical nominal values or online estimates may be used to obtain conservative covariance bounds. As listed in Table~\ref{tab:field_inputs}, the ARW and VRW coefficients are obtained from Allan-variance analysis of approximately 5~h of static raw IMU data, while \(\sigma_v\) is set according to the UM982 receiver specification. The parameters \(\boldsymbol{\epsilon}^b\), \(\boldsymbol{\nabla}^b\), and \(\boldsymbol{l}^b\) are identified on the 600 s dynamic segment using a conventional 18-state SINS/GNSS Kalman filter, which is initialized by the reference baseline and aided by six-dimensional position/velocity measurements. Since the static and dynamic OBA segments are extracted from the same continuous vehicle test, the turn-on biases and mechanically fixed lever arm are common to both segments. The converged steady-state estimates obtained from the dynamically excited segment are therefore used as constant deterministic inputs for both OBA evaluations, as shown in Fig.~\ref{fig:field_overview}(d).

\begin{table}[!t]
\caption{Field-Test Settings and Identified Error Parameters}
\label{tab:field_inputs}
\centering
\footnotesize
\setlength{\tabcolsep}{2pt}
\begin{tabular}{@{}>{\raggedright\arraybackslash}p{0.45\columnwidth}>{\raggedright\arraybackslash}p{0.47\columnwidth}@{}}
\toprule
Parameter & Value \\
\midrule
Sampling rate $f_s$ & $200~\mathrm{Hz}$ \\
Segment duration & $600~\mathrm{s}$ \\
Window length $T_w$ & $10~\mathrm{s}$ \\
Gyroscope bias $\boldsymbol{\epsilon}^b$ & $[0.1879,0.3488,-0.1057]~\mathrm{deg/h}$ \\
Accelerometer bias $\boldsymbol{\nabla}^b$ & $[61.7,-1496.0,-580.8]~\mu\mathrm{g}$ \\
Lever arm $\boldsymbol{l}^b$ & $[0.8041,-0.5724,0.0494]~\mathrm{m}$ \\
GNSS velocity-noise standard deviation $\sigma_v$ & $0.1~\mathrm{m/s}$ \\
Gyroscope angle random walk (ARW) & $[1.856,1.864,0.685]\times10^{-3}~\mathrm{deg}/\sqrt{\mathrm{h}}$ \\
Accelerometer velocity random walk (VRW) & \begin{tabular}[t]{@{}l@{}}
$[8.82,6.44,9.47]\times10^{-3}$ \\
$\mathrm{m/s}/\sqrt{\mathrm{h}}$
\end{tabular} \\
\bottomrule
\end{tabular}
\end{table}

\begin{figure*}[!t]
\centering
\includegraphics[width=0.80\textwidth]{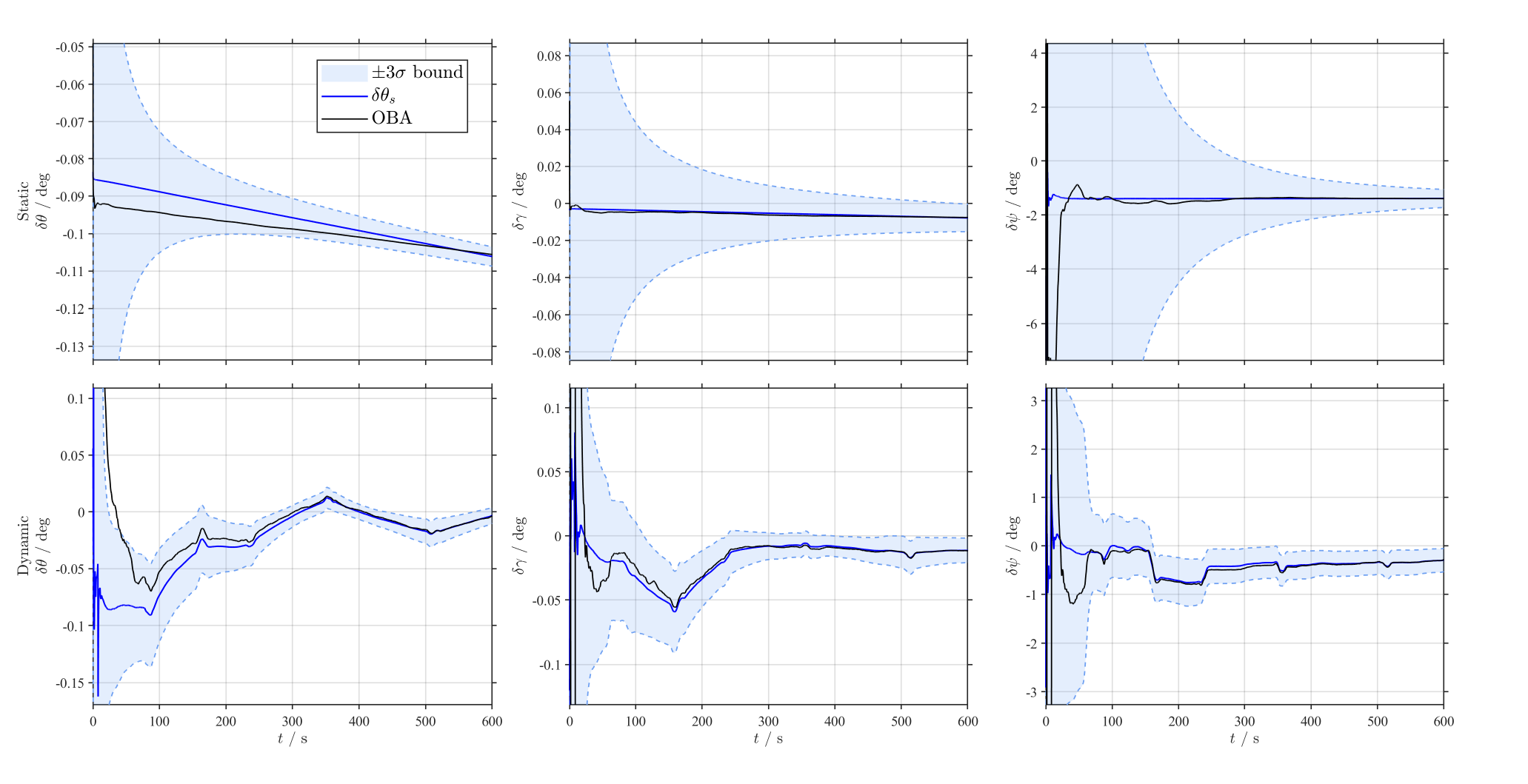}
\caption{Measured OBA initial-attitude errors, predicted offsets, and analytical \(\pm3\sigma\) envelopes for the static and dynamic NS260 field-test segments.}
\label{fig:field_ns260_envelope}
\end{figure*}

After the NS260 OBA initial attitude is obtained, the actual initial-attitude error is computed relative to the FSINS3X/RTK reference baseline. Since the two systems are independently installed, a fixed mounting angle exists between them. The mounting-angle offset is estimated from the mean attitude difference during the final 100 s static interval of the 3200 s test and removed before computing the initial-attitude error. This calibration aligns only the physical reference frames and does not alter the transient error evolution or the statistical spread predicted by the proposed model. Table~\ref{tab:field_result} reports the pitch, roll, and heading components of \(\boldsymbol{r}_{K_{\max}}\) and \(\boldsymbol{r}_{\rm RMSE}\) defined in \eqref{eq:det_residual_metrics}, together with the time-series coverage, i.e., the proportion of epochs at which the measured initial-attitude error lies within the analytical \(\pm3\sigma\) bounds.

\begin{table}[!t]
\caption{Statistical Summary of Field-Test Validation Results}
\label{tab:field_result}
\centering
\footnotesize
\setlength{\tabcolsep}{1.5pt}
\begin{tabular}{llccc}
\toprule
Segment & Axis &
\begin{tabular}[c]{@{}c@{}}Component of\\ \(\boldsymbol{r}_{K_{\max}}\) / deg\end{tabular} &
\begin{tabular}[c]{@{}c@{}}Component of\\ \(\boldsymbol{r}_{\rm RMSE}\) / deg\end{tabular} &
\begin{tabular}[c]{@{}c@{}}Time-series\\coverage\end{tabular} \\
\midrule
Static & Pitch & $5.4\times10^{-4}$ & $3.1\times10^{-4}$ & $0.999$ \\
Static & Roll & $2.2\times10^{-4}$ & $1.3\times10^{-4}$ & $1.000$ \\
Static & Heading & $-5.26\times10^{-3}$ & $3.78\times10^{-3}$ & $0.999$ \\
Dynamic & Pitch & $-3.9\times10^{-4}$ & $4.7\times10^{-4}$ & $0.941$ \\
Dynamic & Roll & $-2.8\times10^{-4}$ & $1.8\times10^{-4}$ & $0.997$ \\
Dynamic & Heading & $1.486\times10^{-2}$ & $4.95\times10^{-3}$ & $0.989$ \\
\bottomrule
\end{tabular}
\end{table}

Fig.~\ref{fig:field_ns260_envelope} compares the measured NS260 initial-attitude errors with the predicted deterministic offsets and analytical \(\pm3\sigma\) envelopes. In the static segment, the heading envelope is significantly wider than the pitch and roll envelopes because the lack of motion excitation makes heading weakly observable. Nevertheless, Table~\ref{tab:field_result} shows that the empirical coverage ratios in the static segment remain close to unity for all three attitude axes, with a maximum component of \(\boldsymbol{r}_{\rm RMSE}\) of \(3.78\times10^{-3}\) deg. In the dynamic segment, vehicle maneuvers enhance observability and accelerate heading covariance convergence. The components of \(\boldsymbol{r}_{K_{\max}}\) remain within \(1.486\times10^{-2}\) deg, those of \(\boldsymbol{r}_{\rm RMSE}\) are below \(4.95\times10^{-3}\) deg, and the empirical coverage ratios remain above \(0.941\). These results verify that the proposed model predicts both deterministic offsets and stochastic envelopes under real vehicle testing.

\section{Conclusion}
\setlength{\parskip}{0ex}
This paper develops a first-order analytical propagation model for GNSS-aided sliding-window velocity-integral OBA coarse alignment. Based on the nominal sliding-window observation model, gyroscope, accelerometer, GNSS-velocity, and lever-arm errors are mapped into perturbations of the unnormalized observation vectors and are then propagated to attitude error through a perturbation analysis of Davenport's q method. The model analytically decouples deterministic attitude-offset prediction from stochastic covariance propagation while accounting for accumulated attitude-projection errors, the difference of the velocity errors at the start and end epochs of each sliding window, and noise correlations induced by overlapping windows.

Simulation results show that the predicted deterministic offsets agree with numerical OBA results, the Monte Carlo standard deviations agree with those predicted by the analytical covariance, and the empirical coverage ratios under simultaneous combined-error injection remain above \(0.994\). Vehicle tests further demonstrate that the NS260 initial-attitude errors are bounded by the predicted offsets and \(\pm3\sigma\) envelopes, with residual RMSE below \(4.95\times10^{-3}\) deg and a maximum final residual of \(1.486\times10^{-2}\) deg. These results indicate that the proposed first-order model provides an effective analytical tool for attitude-accuracy evaluation of GNSS-aided sliding-window OBA coarse alignment. It should be noted that the proposed model evaluates the estimation error of the constant initial attitude, whereas the real-time attitude error at the end of alignment additionally contains continuously accumulated gyroscope integration errors over \([0,t]\). Extending the current framework to real-time attitude error covariance over the full alignment interval remains future work.

\bibliographystyle{IEEEtran}
\bibliography{IEEEabrv,swv_error_propagation_arxiv_v3}

\makeatletter
\def\@IEEEBIOskipN{0pt}
\makeatother

\begin{IEEEbiography}[{\includegraphics[width=1in,height=1.25in,keepaspectratio]{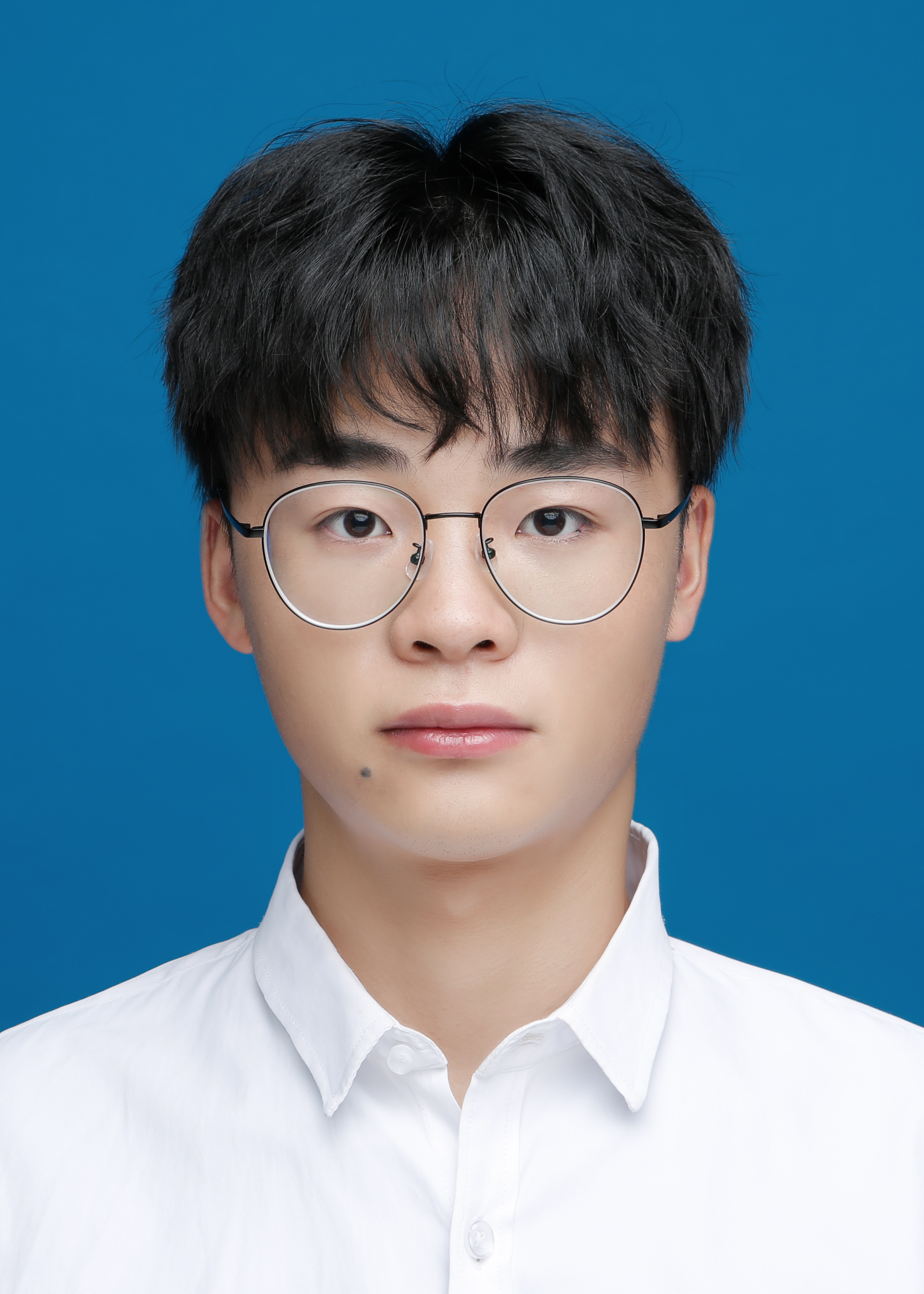}}]{Xuyang Jiang}
(Graduate Student Member, IEEE) received the B.E. degree in measurement and control technology and instrumentation in 2022 from Southeast University, Nanjing, China, where he is currently working toward the Ph.D. degree in instrument science and technology with the School of Instrument Science and Engineering.

His research interests include integrated navigation, signal processing, and information fusion.
\end{IEEEbiography}

\begin{IEEEbiography}[{\includegraphics[width=1in,height=1.25in,keepaspectratio]{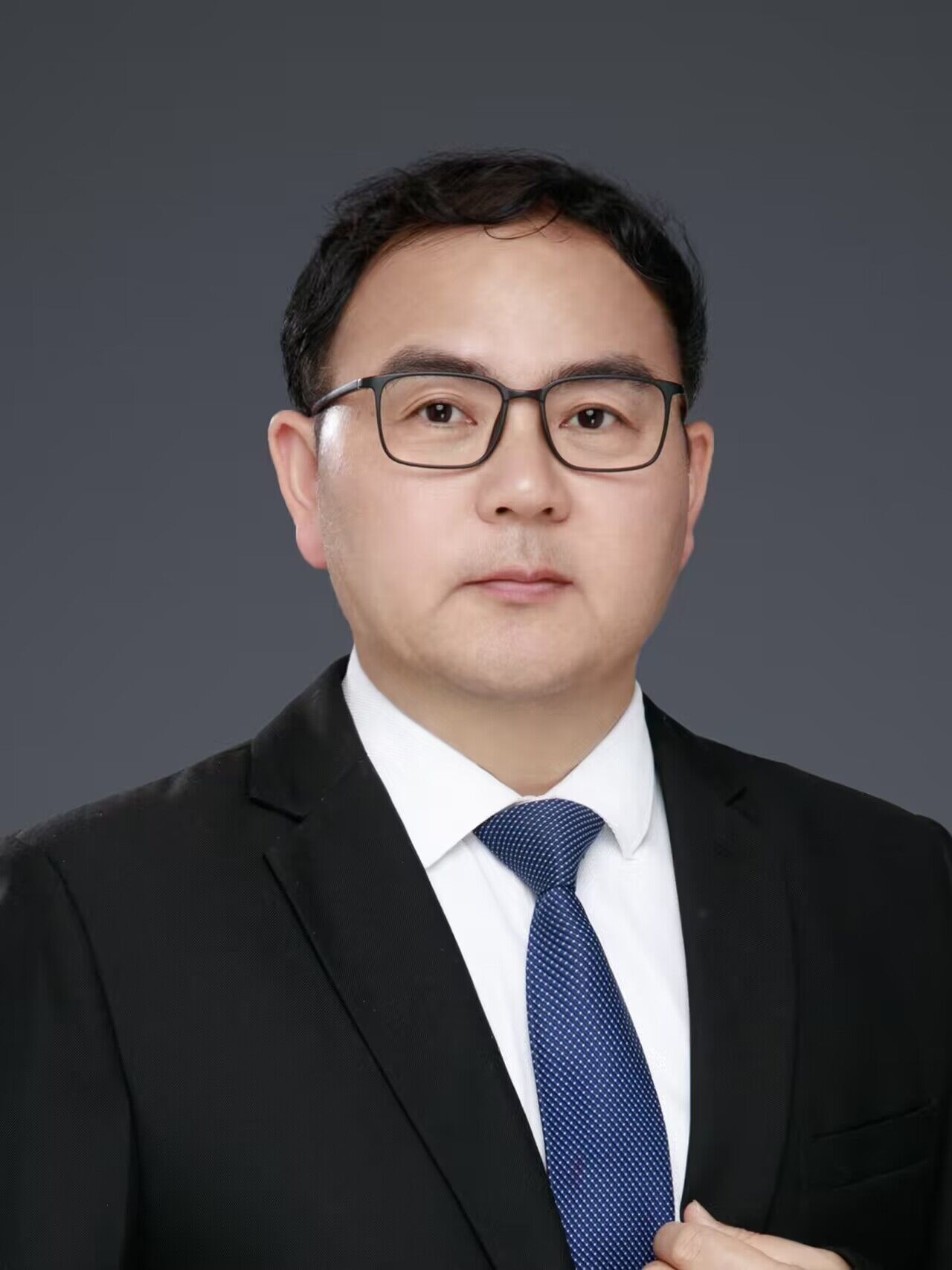}}]{Xiyuan Chen}
(Senior Member, IEEE) received the Ph.D. degree in precision instrument and machinery from Southeast University, Nanjing, China, in 1998.

He is currently a Professor with the School of Instrument Science and Engineering, Southeast University. His research interests include fiber optic sensors, inertial navigation, GNSS software receiver and wireless location technologies, integrated navigation, and related application.
\end{IEEEbiography}

\begin{IEEEbiography}[{\includegraphics[width=1in,height=1.25in,keepaspectratio]{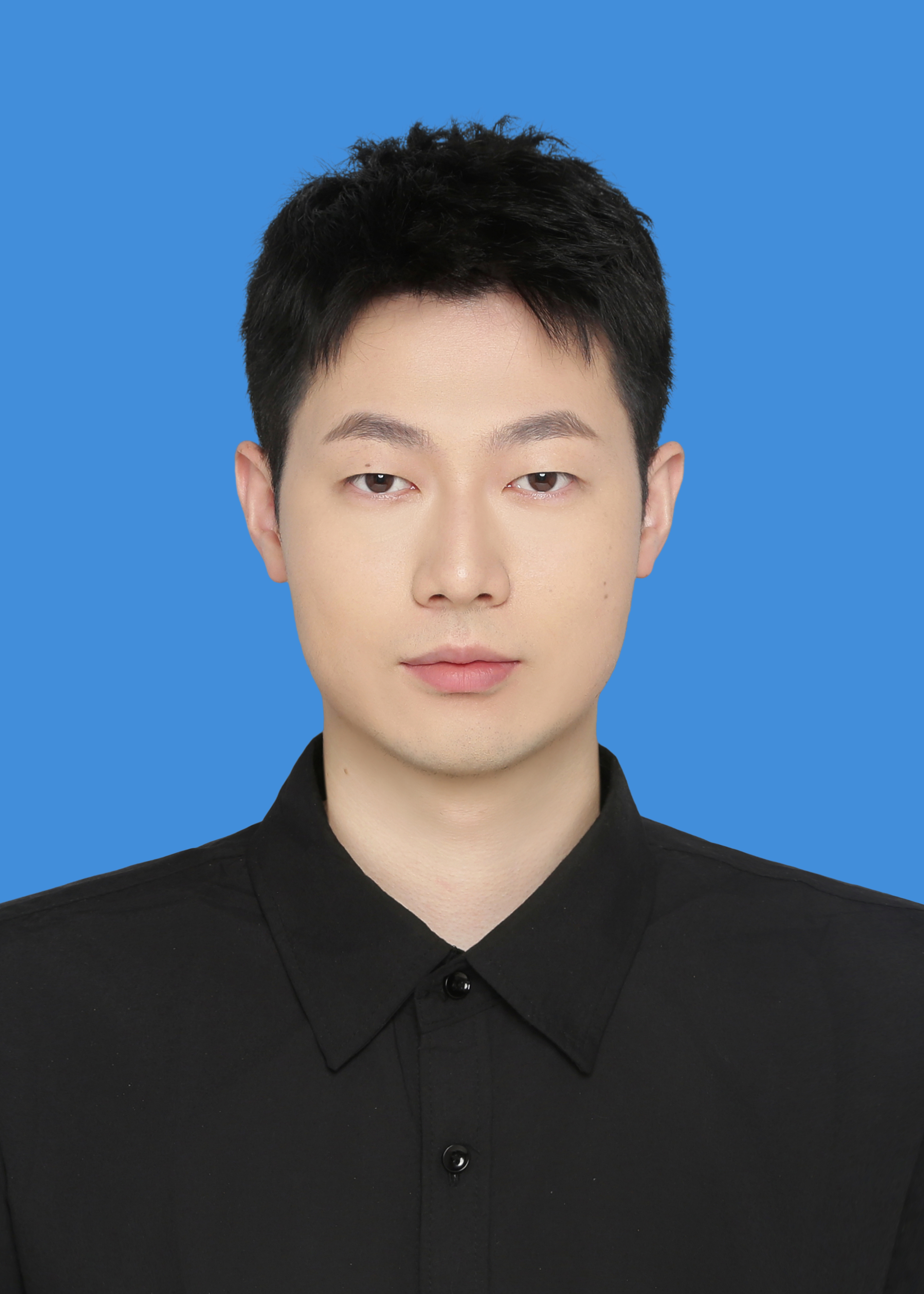}}]{Weiming Jing}
(Graduate Student Member, IEEE) received the B.E. degree in electronic information engineering from China Three Gorges University, Yichang, China, in 2023. He is currently working toward the Eng.D. degree in instrument science and technology with the School of Instrument Science and Engineering, Southeast University, Nanjing, China.

His research interests include 3-D object detection and LLM-based autonomous driving.
\end{IEEEbiography}

\flushcolsend
\clearpage
\raggedcolsend
\raggedend

\makeatletter
\let\maketitle\arxiv@maketitle
\let\@maketitle\arxiv@@maketitle
\let\thanks\arxiv@thanks
\makeatother

\setcounter{section}{0}
\setcounter{subsection}{0}
\setcounter{equation}{0}
\setcounter{footnote}{0}
\renewcommand{\theequation}{S\arabic{equation}}
\renewcommand{\theHequation}{supp.equation.\arabic{equation}}
\renewcommand{\theHsection}{supp.section.\arabic{section}}
\renewcommand{\theHsubsection}{supp.subsection.\arabic{section}.\arabic{subsection}}
\renewcommand{\theHfootnote}{supp.footnote.\arabic{footnote}}
\markboth{}{}

\title{Supplementary Material for ``Optimization-Based Velocity-Integral Sliding-Window Coarse Alignment: Attitude Error Analysis and Validation''}
\author{Xuyang Jiang, \IEEEmembership{Graduate Student Member, IEEE},
Xiyuan Chen, \IEEEmembership{Senior Member, IEEE},
and Weiming Jing, \IEEEmembership{Graduate Student Member, IEEE}
\thanks{This work was supported in part by Guizhou Provincial Key Technology R\&D Program under Grant XKBF [2025] 032, and in part by the National Natural Science Foundation of China under Grant 61873064. (Corresponding author: Xiyuan Chen.)}
\thanks{Xuyang Jiang and Weiming Jing are with the School of Instrument Science and Engineering, State Key Laboratory of comprehensive PNT Network and Equipment Technology, Southeast University, Nanjing 210096, China, and also with the Key Laboratory of MicroInertial Instrument and Advanced Navigation Technology of Ministry of Education, Southeast University, Nanjing 210096, China (e-mail: jxy@seu.edu.cn; jwm@seu.edu.cn).}
\thanks{Xiyuan Chen is with the School of Instrument Science and Engineering, State Key Laboratory of comprehensive PNT Network and Equipment Technology, Southeast University, Nanjing 210096, China, also with the Key Laboratory of Micro-Inertial Instrument and Advanced Navigation Technology of Ministry of Education, Southeast University, Nanjing 210096, China, and also with the School of Aerospace Engineering, Guizhou Institute of Technology, Guizhou 550025, China (e-mail: chxiyuan@seu.edu.cn).}}
\maketitle

Unless otherwise stated, all symbols follow the definitions in the main paper.

\section{Continuous Observation Model Derivation}
\label{sec:supp_continuous}

This section derives the body- and navigation-side velocity-integral observation vectors in (4)--(6) of the main paper. The derivation begins with the specific-force equation in (3) of the main paper:
\begin{equation}
\dot{\boldsymbol{v}}^n
=\mathbf{C}_b^n\boldsymbol{f}^b
-\left(2\boldsymbol{\omega}_{ie}^n+\boldsymbol{\omega}_{en}^n\right)
\times\boldsymbol{v}^n+\boldsymbol{g}^n .
\label{eq:supp_vel_eq}
\end{equation}
Using the attitude-chain decomposition in (1) of the main paper gives
\begin{equation}
\mathbf{C}_b^n(t)=\mathbf{C}_{n(0)}^{n(t)}\mathbf{C}_b^n(0)\mathbf{C}_{b(t)}^{b(0)} .
\label{eq:supp_chain}
\end{equation}
The navigation-frame rotation rate satisfies
\begin{equation}
\boldsymbol{\omega}_{in}^n
=\boldsymbol{\omega}_{ie}^n+\boldsymbol{\omega}_{en}^n ,
\label{eq:supp_win}
\end{equation}
where, under the ENU convention,
\begin{equation}
\boldsymbol{\omega}_{ie}^n=
\begin{bmatrix}
0\\ \omega_e\cos L\\ \omega_e\sin L
\end{bmatrix},
\quad
\boldsymbol{\omega}_{en}^n=
\begin{bmatrix}
-v_N/(R_M+h)\\
v_E/(R_N+h)\\
v_E\tan L/(R_N+h)
\end{bmatrix}.
\label{eq:supp_rates}
\end{equation}
Here \(\omega_e\) is the Earth rotation rate, \(L\) is latitude, \(h\) is height, and \(R_M\) and \(R_N\) are the meridian and transverse radii of curvature, respectively.
Substituting \eqref{eq:supp_chain} into \eqref{eq:supp_vel_eq} and left-multiplying by \(\mathbf{C}_{n(t)}^{n(0)}\) give
\begin{equation}
\begin{aligned}
\mathbf{C}_b^n(0)\mathbf{C}_{b(t)}^{b(0)}\boldsymbol{f}^b
&=\mathbf{C}_{n(t)}^{n(0)}
\left[
\dot{\boldsymbol{v}}^n
+(2\boldsymbol{\omega}_{ie}^n+\boldsymbol{\omega}_{en}^n)
\times\boldsymbol{v}^n
-\boldsymbol{g}^n
\right].
\end{aligned}
\label{eq:supp_obs_inst}
\end{equation}
Integrating \eqref{eq:supp_obs_inst} over \([t_m,t]\) yields
\begin{equation}
\mathbf{C}_b^n(0)\boldsymbol{\alpha}(t)=\boldsymbol{\beta}(t),
\label{eq:supp_obs}
\end{equation}
with
\begin{equation}
\boldsymbol{\alpha}(t)=
\int_{t_m}^{t}\mathbf{C}_{b(\tau)}^{b(0)}\boldsymbol{f}^b(\tau)d\tau .
\label{eq:supp_alpha}
\end{equation}
To obtain the navigation-side vector, use
\begin{equation}
\frac{d}{d\tau}
\left[\mathbf{C}_{n(\tau)}^{n(0)}\boldsymbol{v}^n(\tau)\right]
=\mathbf{C}_{n(\tau)}^{n(0)}
\left[
\dot{\boldsymbol{v}}^n(\tau)
+\boldsymbol{\omega}_{in}^n(\tau)\times\boldsymbol{v}^n(\tau)
\right].
\label{eq:supp_parts}
\end{equation}
Thus,
\begin{equation}
\begin{aligned}
\int_{t_m}^{t}\mathbf{C}_{n(\tau)}^{n(0)}
\dot{\boldsymbol{v}}^n(\tau)d\tau
&=\mathbf{C}_{n(t)}^{n(0)}\boldsymbol{v}^n(t)
-\mathbf{C}_{n(t_m)}^{n(0)}\boldsymbol{v}^n(t_m)\\
&\quad-
\int_{t_m}^{t}\mathbf{C}_{n(\tau)}^{n(0)}
\left[
\boldsymbol{\omega}_{in}^n(\tau)\times\boldsymbol{v}^n(\tau)
\right]d\tau .
\end{aligned}
\label{eq:supp_int_parts}
\end{equation}
Since
\((2\boldsymbol{\omega}_{ie}^n+\boldsymbol{\omega}_{en}^n)
-\boldsymbol{\omega}_{in}^n
=\boldsymbol{\omega}_{ie}^n\),
the navigation-side vector becomes
\begin{equation}
\begin{aligned}
\boldsymbol{\beta}(t)
&=\mathbf{C}_{n(t)}^{n(0)}\boldsymbol{v}^n(t)
-\mathbf{C}_{n(t_m)}^{n(0)}\boldsymbol{v}^n(t_m)\\
&\quad+
\int_{t_m}^{t}\mathbf{C}_{n(\tau)}^{n(0)}
\left[
\boldsymbol{\omega}_{ie}^n(\tau)\times\boldsymbol{v}^n(\tau)
-\boldsymbol{g}^n(\tau)
\right]d\tau .
\end{aligned}
\label{eq:supp_beta}
\end{equation}

\section{Discrete Attitude-Chain Propagation and Specific-Force Increment}
\label{sec:supp_discrete}

This section derives (7)--(10) of the main paper. The coning-compensated body-frame rotation vector is
\begin{equation}
\boldsymbol{\phi}_{b,k}
=\Delta\boldsymbol{\theta}_{1,k}
+\Delta\boldsymbol{\theta}_{2,k}
+\frac{2}{3}\Delta\boldsymbol{\theta}_{1,k}
\times\Delta\boldsymbol{\theta}_{2,k}.
\label{eq:supp_coning}
\end{equation}
With \(\phi_{b,k}=\|\boldsymbol{\phi}_{b,k}\|\), the single-step attitude increment is computed by Rodrigues' formula:
\begin{equation}
\begin{aligned}
\mathbf{C}_{b(t_{k+1})}^{b(t_k)}
&=\mathbf{I}_3+
\frac{\sin\phi_{b,k}}{\phi_{b,k}}\crossmat{\boldsymbol{\phi}_{b,k}}\\
&\quad+
\frac{1-\cos\phi_{b,k}}{\phi_{b,k}^2}
\crossmat{\boldsymbol{\phi}_{b,k}}^2 .
\end{aligned}
\label{eq:supp_rodrigues}
\end{equation}
The accumulated body attitude is propagated by
\begin{equation}
\mathbf{C}_{b(t_{k+1})}^{b(0)}
=\mathbf{C}_{b(t_k)}^{b(0)}\mathbf{C}_{b(t_{k+1})}^{b(t_k)} .
\label{eq:supp_body_prop}
\end{equation}
The navigation-frame attitude chain is propagated analogously using the velocity-derived angular rate \(\boldsymbol{\omega}_{in}^n\):
\begin{equation}
\mathbf{C}_{n(t_{k+1})}^{n(0)}
=\mathbf{C}_{n(t_k)}^{n(0)}\mathbf{C}_{n(t_{k+1})}^{n(t_k)} .
\label{eq:supp_nav_prop}
\end{equation}
The incremental matrix \(\mathbf{C}_{n(t_{k+1})}^{n(t_k)}\) can be computed with the same Rodrigues form as \eqref{eq:supp_rodrigues} after replacing \(\boldsymbol{\phi}_{b,k}\) by the navigation-frame rotation vector over \([t_k,t_{k+1}]\).

The body-side contribution over one update period is
\begin{equation}
\Delta\boldsymbol{\alpha}_k
=\mathbf{C}_{b(t_k)}^{b(0)}
\int_0^T \mathbf{C}_{b(t_k+s)}^{b(t_k)}\boldsymbol{f}^b(s)ds .
\label{eq:supp_alpha_step}
\end{equation}
The dual-sample increments are defined by
\begin{equation}
\begin{aligned}
\Delta\boldsymbol{\theta}_{1,k}&=\int_0^{T/2}\boldsymbol{\omega}^b(s)ds,&
\Delta\boldsymbol{\theta}_{2,k}&=\int_{T/2}^{T}\boldsymbol{\omega}^b(s)ds,\\
\Delta\boldsymbol{v}_{1,k}&=\int_0^{T/2}\boldsymbol{f}^b(s)ds,&
\Delta\boldsymbol{v}_{2,k}&=\int_{T/2}^{T}\boldsymbol{f}^b(s)ds.
\end{aligned}
\label{eq:supp_half_increments}
\end{equation}
Assume that angular rate and specific force vary linearly over the update period:
\begin{equation}
\boldsymbol{\omega}^b(s)=\boldsymbol{a}+\boldsymbol{b}s,
\quad
\boldsymbol{f}^b(s)=\boldsymbol{c}+\boldsymbol{d}s,
\quad 0\le s\le T .
\label{eq:supp_linear}
\end{equation}
Here \(s\) is the local time within \([t_k,t_{k+1}]\), and \(\boldsymbol{a}\), \(\boldsymbol{b}\), \(\boldsymbol{c}\), and \(\boldsymbol{d}\) are the linear-model coefficients determined by the two half-sample increments:
\begin{equation}
\boldsymbol{a}=\frac{3\Delta\boldsymbol{\theta}_{1,k}
-\Delta\boldsymbol{\theta}_{2,k}}{T},
\quad
\boldsymbol{b}=\frac{4(\Delta\boldsymbol{\theta}_{2,k}
-\Delta\boldsymbol{\theta}_{1,k})}{T^2},
\label{eq:supp_ab}
\end{equation}
\begin{equation}
\boldsymbol{c}=\frac{3\Delta\boldsymbol{v}_{1,k}
-\Delta\boldsymbol{v}_{2,k}}{T},
\quad
\boldsymbol{d}=\frac{4(\Delta\boldsymbol{v}_{2,k}
-\Delta\boldsymbol{v}_{1,k})}{T^2}.
\label{eq:supp_cd}
\end{equation}
To derive the rotation and sculling correction terms, the attitude-chain matrix over a short interval is expanded to first order as
\begin{equation}
\mathbf{C}_{b(t_k+s)}^{b(t_k)}
\approx \mathbf{I}_3+
\left[
\int_0^s\boldsymbol{\omega}^b(\lambda)d\lambda
\times
\right].
\label{eq:supp_first_order}
\end{equation}
With \eqref{eq:supp_first_order}, the integral in \eqref{eq:supp_alpha_step} separates as
\begin{equation}
\begin{aligned}
\int_0^T \mathbf{C}_{b(t_k+s)}^{b(t_k)}\boldsymbol{f}^b(s)ds
&\approx \int_0^T\boldsymbol{f}^b(s)ds\\
&\quad+\int_0^T\left(\int_0^s\boldsymbol{\omega}^b(\lambda)d\lambda\right)
\times\boldsymbol{f}^b(s)ds .
\end{aligned}
\label{eq:supp_split_integral}
\end{equation}
The first term is \(\Delta\boldsymbol{v}_{1,k}+\Delta\boldsymbol{v}_{2,k}\). Substitution of \eqref{eq:supp_linear}--\eqref{eq:supp_cd} into the second term gives the overall rotation correction
\(\frac12(\Delta\boldsymbol{\theta}_{1,k}+\Delta\boldsymbol{\theta}_{2,k})\times(\Delta\boldsymbol{v}_{1,k}+\Delta\boldsymbol{v}_{2,k})\)
and the dual-sample sculling correction
\(\frac23(\Delta\boldsymbol{\theta}_{1,k}\times\Delta\boldsymbol{v}_{2,k}+\Delta\boldsymbol{v}_{1,k}\times\Delta\boldsymbol{\theta}_{2,k})\). Therefore,
\begin{equation}
\begin{aligned}
\Delta\boldsymbol{v}_{c,k}
&=\Delta\boldsymbol{v}_{1,k}+\Delta\boldsymbol{v}_{2,k}\\
&\quad+\frac{1}{2}
(\Delta\boldsymbol{\theta}_{1,k}+\Delta\boldsymbol{\theta}_{2,k})
\times
(\Delta\boldsymbol{v}_{1,k}+\Delta\boldsymbol{v}_{2,k})\\
&\quad+\frac{2}{3}
\left(
\Delta\boldsymbol{\theta}_{1,k}\times\Delta\boldsymbol{v}_{2,k}
+\Delta\boldsymbol{v}_{1,k}\times\Delta\boldsymbol{\theta}_{2,k}
\right).
\end{aligned}
\label{eq:supp_dvc}
\end{equation}

\section{Residual-Minimization OBA Matrix and Davenport Gain Matrix Equivalence}
\label{sec:supp_equivalence}

Given \(N\) observation pairs, the OBA attitude estimate is obtained by minimizing the weighted sum of squared residuals:
\begin{equation}
J(\boldsymbol{q})=\sum_{i=1}^{N}w_i
\left\|
\boldsymbol{\beta}_i-\mathbf{C}(\boldsymbol{q})\boldsymbol{\alpha}_i
\right\|^2 ,
\label{eq:supp_residual_cost}
\end{equation}
where \(\mathbf{C}(\boldsymbol{q})\) is the attitude matrix corresponding to \(\boldsymbol{q}\). Expanding \eqref{eq:supp_residual_cost} gives
\begin{equation}
J(\boldsymbol{q})=
\sum_{i=1}^{N}w_i
\left(
\|\boldsymbol{\alpha}_i\|^2+\|\boldsymbol{\beta}_i\|^2
\right)
-2\sum_{i=1}^{N}w_i\boldsymbol{\beta}_i^{\mathrm{T}}\mathbf{C}(\boldsymbol{q})\boldsymbol{\alpha}_i .
\label{eq:supp_residual_expand}
\end{equation}
The Davenport gain matrix \(\mathbf{K}\) defined in the main paper satisfies
\begin{equation}
\boldsymbol{q}^{\mathrm{T}}\mathbf{K}\boldsymbol{q}
=\sum_{i=1}^{N}w_i\boldsymbol{\beta}_i^{\mathrm{T}}\mathbf{C}(\boldsymbol{q})\boldsymbol{\alpha}_i .
\label{eq:supp_qKq}
\end{equation}
Using \(\boldsymbol{q}^{\mathrm{T}}\boldsymbol{q}=1\), the weighted least-squares cost \(J(\boldsymbol{q})=c-2\boldsymbol{q}^{\mathrm{T}}\mathbf{K}\boldsymbol{q}\) can be written as the quadratic form
\begin{equation}
J(\boldsymbol{q})=\boldsymbol{q}^{\mathrm{T}}\mathbf{K}_{\mathrm{OBA}}\boldsymbol{q},
\quad
\mathbf{K}_{\mathrm{OBA}}=c\mathbf{I}_4-2\mathbf{K},
\label{eq:supp_Koba}
\end{equation}
where
\begin{equation}
c=\sum_{i=1}^{N}w_i
\left(
\|\boldsymbol{\alpha}_i\|^2+\|\boldsymbol{\beta}_i\|^2
\right).
\label{eq:supp_c}
\end{equation}
Since \(c\mathbf{I}_4\) does not change the eigenvectors and the factor \(-2\) reverses the eigenvalue ordering, the minimum-eigenvalue eigenvector of \(\mathbf{K}_{\mathrm{OBA}}\) is the maximum-eigenvalue eigenvector of \(\mathbf{K}\), thereby yielding the same attitude quaternion solution.

\section{Projection-Error Model}
\label{sec:supp_projection}

For either rotating frame \(x\in\{b,n\}\), the true and computed attitude-chain matrices are propagated by
\begin{equation}
\begin{aligned}
\dot{\mathbf{C}}_{x(\tau)}^{x(0)}
&=\mathbf{C}_{x(\tau)}^{x(0)}\crossmat{\boldsymbol{\omega}^x(\tau)},\\
\dot{\widetilde{\mathbf{C}}}_{x(\tau)}^{x(0)}
&=\widetilde{\mathbf{C}}_{x(\tau)}^{x(0)}
\crossmat{\boldsymbol{\omega}^x(\tau)+\delta\boldsymbol{\omega}^x(\tau)} .
\end{aligned}
\label{eq:supp_chain_prop_error}
\end{equation}
Under a right-multiplicative attitude-error model, the computed matrix is expressed as
\begin{equation}
\widetilde{\mathbf{C}}_{x(\tau)}^{x(0)}
\approx \mathbf{C}_{x(\tau)}^{x(0)}
\left(\mathbf{I}_3+\crossmat{\boldsymbol{\phi}^x(\tau)}\right),
\label{eq:supp_right_error}
\end{equation}
where \(x=b\) and \(x=n\) correspond to the body and navigation rotating frames, respectively. Differentiating \eqref{eq:supp_right_error} and substituting \eqref{eq:supp_chain_prop_error} yield
\begin{equation}
\begin{aligned}
\crossmat{\boldsymbol{\omega}^x}\crossmat{\boldsymbol{\phi}^x}
+\crossmat{\dot{\boldsymbol{\phi}}^x}
&=\crossmat{\delta\boldsymbol{\omega}^x}
+\crossmat{\boldsymbol{\phi}^x}\crossmat{\boldsymbol{\omega}^x}.
\end{aligned}
\label{eq:supp_phi_matrix_balance}
\end{equation}
The second-order product
\(\crossmat{\boldsymbol{\phi}^x}\crossmat{\delta\boldsymbol{\omega}^x}\)
is neglected. Using the commutator identity
\begin{equation}
\crossmat{\boldsymbol{\omega}^x}\crossmat{\boldsymbol{\phi}^x}
-\crossmat{\boldsymbol{\phi}^x}\crossmat{\boldsymbol{\omega}^x}
=\crossmat{\boldsymbol{\omega}^x\times\boldsymbol{\phi}^x}
\label{eq:supp_commutator}
\end{equation}
give
\begin{equation}
\dot{\boldsymbol{\phi}}^x
=\delta\boldsymbol{\omega}^x
-\boldsymbol{\omega}^x\times\boldsymbol{\phi}^x,
\label{eq:supp_phi_dot}
\end{equation}
whose integral solution is
\begin{equation}
\boldsymbol{\phi}^x(\tau)
=\int_0^\tau
\mathbf{C}_{x(s)}^{x(\tau)}
\delta\boldsymbol{\omega}^x(s)ds .
\label{eq:supp_phi}
\end{equation}
The projection error in \eqref{eq:supp_phi} accumulates from the initial alignment epoch. Therefore, even though each observation vector is integrated only over a fixed sliding window, the attitude-chain projection error within that window retains contributions from gyroscope errors occurring before the window start.

\section{Observation-Vector Error Propagation}
\label{sec:supp_observation_error}

This section derives (19)--(22) of the main paper in continuous form.

\subsection{Body-Side Perturbation}
\label{sec:supp_body_error}

For the body-side attitude chain, setting \(x=b\) and \(\delta\boldsymbol{\omega}^x=\delta\boldsymbol{\omega}_{ib}^b\) in \eqref{eq:supp_phi} gives
\begin{equation}
\boldsymbol{\phi}^b(\tau)
=\int_0^\tau
\mathbf{C}_{b(s)}^{b(\tau)}
\delta\boldsymbol{\omega}_{ib}^b(s)\,ds .
\label{eq:supp_phi_body}
\end{equation}
Thus, each past gyroscope error is first projected into the current body frame before being accumulated. The perturbed body-side observation vector is
\begin{equation}
\widetilde{\boldsymbol{\alpha}}_i
=\int_{t_{m_i}}^{t_{M_i}}
\widetilde{\mathbf{C}}_{b(\tau)}^{b(0)}
\widetilde{\boldsymbol{f}}^b(\tau)d\tau ,
\label{eq:supp_alpha_tilde}
\end{equation}
where
\begin{equation}
\widetilde{\mathbf{C}}_{b(\tau)}^{b(0)}
\approx
\mathbf{C}_{b(\tau)}^{b(0)}
\left(\mathbf{I}_3+\crossmat{\boldsymbol{\phi}^b(\tau)}\right),
\quad
\widetilde{\boldsymbol{f}}^b
=\boldsymbol{f}^b+\delta\boldsymbol{f}^b .
\label{eq:supp_alpha_tilde_parts}
\end{equation}
Neglecting the second-order product
\(\crossmat{\boldsymbol{\phi}^b}\delta\boldsymbol{f}^b\), the first-order expansion becomes
\begin{equation}
\delta\boldsymbol{\alpha}_i
=\int_{t_{m_i}}^{t_{M_i}}
\mathbf{C}_{b(\tau)}^{b(0)}
\left[
\delta\boldsymbol{f}^b(\tau)
+\boldsymbol{\phi}^b(\tau)\times\boldsymbol{f}^b(\tau)
\right]d\tau .
\label{eq:supp_alpha_expand}
\end{equation}
Using \(\boldsymbol{\phi}^b\times\boldsymbol{f}^b=-\boldsymbol{f}^b\times\boldsymbol{\phi}^b\), the body-side perturbation is separated into the direct accelerometer-error contribution
\begin{equation}
\delta\boldsymbol{\alpha}_{a,i}
=\int_{t_{m_i}}^{t_{M_i}}
\mathbf{C}_{b(\tau)}^{b(0)}
\delta\boldsymbol{f}^b(\tau)d\tau ,
\label{eq:supp_alpha_a}
\end{equation}
and the gyroscope-error-induced projection contribution
\begin{equation}
\delta\boldsymbol{\alpha}_{g,i}
=-\int_{t_{m_i}}^{t_{M_i}}
\mathbf{C}_{b(\tau)}^{b(0)}
\left[
\boldsymbol{f}^b(\tau)\times\boldsymbol{\phi}^b(\tau)
\right]d\tau .
\label{eq:supp_alpha_g}
\end{equation}
Consequently,
\begin{equation}
\widetilde{\boldsymbol{\alpha}}_i-\boldsymbol{\alpha}_i
=\delta\boldsymbol{\alpha}_{a,i}
+\delta\boldsymbol{\alpha}_{g,i}.
\label{eq:supp_alpha_total}
\end{equation}

\subsection{Navigation-Side Perturbation and Lever-Arm Transfer}
\label{sec:supp_navigation_error}

The velocity used to construct the navigation-side observation comprises the nominal velocity, the GNSS velocity error, and the lever-arm-induced velocity offset:
\begin{equation}
\widetilde{\boldsymbol{v}}^n(\tau)
=\boldsymbol{v}^n(\tau)
+\delta\boldsymbol{v}_{\rm GNSS}^n(\tau)
+\mathbf{C}_{b(\tau)}^{n(\tau)}
\left[
\boldsymbol{\omega}_{eb}^b(\tau)\times\boldsymbol{l}^b
\right].
\label{eq:supp_velocity_perturbed}
\end{equation}
With the Earth-rotation rate treated as known and position errors neglected, the navigation-frame angular-rate perturbation caused by the GNSS velocity error is
\begin{equation}
\begin{aligned}
\delta\boldsymbol{\omega}_{in}^n
&\approx\delta\boldsymbol{\omega}_{en}^n
=\mathbf{M}_v\delta\boldsymbol{v}_{\rm GNSS}^n,\\
\mathbf{M}_v
&=
\begin{bmatrix}
0 & -\dfrac{1}{R_M+h} & 0\\
\dfrac{1}{R_N+h} & 0 & 0\\
\dfrac{\tan L}{R_N+h} & 0 & 0
\end{bmatrix}.
\end{aligned}
\label{eq:supp_transport_rate_error}
\end{equation}
Here, \(R_M\) and \(R_N\) are the meridian and transverse radii of curvature, respectively, \(L\) is latitude, and \(h\) is height. Substitution into \eqref{eq:supp_phi} gives
\begin{equation}
\boldsymbol{\phi}^n(\tau)
=\int_0^\tau
\mathbf{C}_{n(s)}^{n(\tau)}
\mathbf{M}_v(s)
\delta\boldsymbol{v}_{\rm GNSS}^n(s)\,ds .
\label{eq:supp_phi_navigation}
\end{equation}

The navigation-side observation constructed from the perturbed velocity and attitude chain is
\begin{equation}
\begin{aligned}
\widetilde{\boldsymbol{\beta}}_i
={}&\widetilde{\mathbf{C}}_{n(t_{M_i})}^{n(0)}
\widetilde{\boldsymbol{v}}^n(t_{M_i})
-\widetilde{\mathbf{C}}_{n(t_{m_i})}^{n(0)}
\widetilde{\boldsymbol{v}}^n(t_{m_i})\\
&+\int_{t_{m_i}}^{t_{M_i}}
\widetilde{\mathbf{C}}_{n(\tau)}^{n(0)}
\left[
\boldsymbol{\omega}_{ie}^n(\tau)\times
\widetilde{\boldsymbol{v}}^n(\tau)
-\boldsymbol{g}^n(\tau)
\right]d\tau .
\end{aligned}
\label{eq:supp_beta_perturbed}
\end{equation}
Using \eqref{eq:supp_right_error} with \(x=n\) and retaining only first-order terms gives
\begin{equation}
\begin{aligned}
\widetilde{\boldsymbol{\beta}}_i-\boldsymbol{\beta}_i
={}&\mathbf{C}_{n(t_{M_i})}^{n(0)}
\delta\boldsymbol{v}_{\rm GNSS}^n(t_{M_i})
-\mathbf{C}_{n(t_{m_i})}^{n(0)}
\delta\boldsymbol{v}_{\rm GNSS}^n(t_{m_i})\\
&+\mathbf{C}_{n(t_{M_i})}^{n(0)}
\crossmat{\boldsymbol{\phi}^n(t_{M_i})}
\boldsymbol{v}^n(t_{M_i})\\
&-\mathbf{C}_{n(t_{m_i})}^{n(0)}
\crossmat{\boldsymbol{\phi}^n(t_{m_i})}
\boldsymbol{v}^n(t_{m_i})\\
&+\int_{t_{m_i}}^{t_{M_i}}
\mathbf{C}_{n(\tau)}^{n(0)}
\boldsymbol{\omega}_{ie}^n(\tau)\times
\delta\boldsymbol{v}_{\rm GNSS}^n(\tau)d\tau\\
&+\int_{t_{m_i}}^{t_{M_i}}
\mathbf{C}_{n(\tau)}^{n(0)}
\boldsymbol{\phi}^n(\tau)\times
\left[\boldsymbol{\omega}_{ie}^n(\tau)\times\boldsymbol{v}^n(\tau)\right]
d\tau\\
&-\int_{t_{m_i}}^{t_{M_i}}
\mathbf{C}_{n(\tau)}^{n(0)}
\left[\boldsymbol{\phi}^n(\tau)\times\boldsymbol{g}^n(\tau)\right]
d\tau\\
&+\mathbf{C}_{n(t_{M_i})}^{n(0)}
\Bigl\{\mathbf{C}_{b(t_{M_i})}^{n(t_{M_i})}
\left[
\boldsymbol{\omega}_{eb}^b(t_{M_i})\times\boldsymbol{l}^b
\right]\Bigr\}\\
&-\mathbf{C}_{n(t_{m_i})}^{n(0)}
\Bigl\{\mathbf{C}_{b(t_{m_i})}^{n(t_{m_i})}
\left[
\boldsymbol{\omega}_{eb}^b(t_{m_i})\times\boldsymbol{l}^b
\right]\Bigr\}\\
&+\int_{t_{m_i}}^{t_{M_i}}
\mathbf{C}_{n(\tau)}^{n(0)}
\boldsymbol{\omega}_{ie}^n(\tau)\times
{}\\
&\qquad
\mathbf{C}_{b(\tau)}^{n(\tau)}
\left[
\boldsymbol{\omega}_{eb}^b(\tau)\times\boldsymbol{l}^b
\right]
d\tau .
\end{aligned}
\label{eq:supp_beta_error_full}
\end{equation}
After neglecting second-order products, the retained GNSS-related terms are
\begin{equation}
\begin{aligned}
\delta\boldsymbol{\beta}_{v,i}
={}&\mathbf{C}_{n(t_{M_i})}^{n(0)}
\delta\boldsymbol{v}_{\rm GNSS}^n(t_{M_i})
-\mathbf{C}_{n(t_{m_i})}^{n(0)}
\delta\boldsymbol{v}_{\rm GNSS}^n(t_{m_i})\\
&+\mathbf{C}_{n(t_{M_i})}^{n(0)}
\crossmat{\boldsymbol{\phi}^n(t_{M_i})}
\boldsymbol{v}^n(t_{M_i})\\
&-\mathbf{C}_{n(t_{m_i})}^{n(0)}
\crossmat{\boldsymbol{\phi}^n(t_{m_i})}
\boldsymbol{v}^n(t_{m_i})\\
&+\int_{t_{m_i}}^{t_{M_i}}
\mathbf{C}_{n(\tau)}^{n(0)}
\Bigl\{
\boldsymbol{\omega}_{ie}^n(\tau)\times
\delta\boldsymbol{v}_{\rm GNSS}^n(\tau)\\
&\qquad\qquad+
\boldsymbol{\phi}^n(\tau)\times
\left[
\boldsymbol{\omega}_{ie}^n(\tau)\times\boldsymbol{v}^n(\tau)
-\boldsymbol{g}^n(\tau)
\right]
\Bigr\}d\tau .
\end{aligned}
\label{eq:supp_beta_v_full}
\end{equation}

Using the attitude-chain identity
\begin{equation}
\mathbf{C}_{n(\tau)}^{n(0)}\mathbf{C}_{b(\tau)}^{n(\tau)}
=\mathbf{C}_b^n(0)\mathbf{C}_{b(\tau)}^{b(0)}
\label{eq:supp_chain_identity}
\end{equation}
and the rotational invariance of the cross product, the lever-arm terms can be transferred directly to the body side as \(\mathbf{C}_b^n(0)\delta\boldsymbol{\alpha}_{l,i}\), where
\begin{equation}
\begin{aligned}
\delta\boldsymbol{\alpha}_{l,i}
={}&\mathbf{C}_{b(t_{M_i})}^{b(0)}
\left[
\boldsymbol{\omega}_{eb}^b(t_{M_i})\times\boldsymbol{l}^b
\right]\\
&-\mathbf{C}_{b(t_{m_i})}^{b(0)}
\left[
\boldsymbol{\omega}_{eb}^b(t_{m_i})\times\boldsymbol{l}^b
\right]\\
&+\int_{t_{m_i}}^{t_{M_i}}
\mathbf{C}_{b(\tau)}^{b(0)}
\left\{
\boldsymbol{\omega}_{ie}^b(\tau)\times
\left[
\boldsymbol{\omega}_{eb}^b(\tau)\times\boldsymbol{l}^b
\right]
\right\}d\tau .
\end{aligned}
\label{eq:supp_alpha_lever_full}
\end{equation}
Thus,
\begin{equation}
\widetilde{\boldsymbol{\beta}}_i-\boldsymbol{\beta}_i
=\delta\boldsymbol{\beta}_{v,i}
+\mathbf{C}_b^n(0)\delta\boldsymbol{\alpha}_{l,i}.
\label{eq:supp_beta_error_decomposition}
\end{equation}
The equivalent observation-vector perturbations entering Davenport's q method are therefore
\begin{equation}
\delta\boldsymbol{\alpha}_i
=\delta\boldsymbol{\alpha}_{a,i}
+\delta\boldsymbol{\alpha}_{g,i}
-\delta\boldsymbol{\alpha}_{l,i},
\qquad
\delta\boldsymbol{\beta}_i
=\delta\boldsymbol{\beta}_{v,i}.
\label{eq:supp_equivalent_observation_errors}
\end{equation}

\section{Davenport's q Method Perturbation and Observation-Vector Jacobians}
\label{sec:supp_davenport}

\subsection{Eigenvector Perturbation}
\label{sec:supp_eigen}

The first-order eigenvector perturbation required to derive \(\boldsymbol{\Phi}\) is obtained by linearizing the perturbed eigenvalue equation in Davenport's q method:
\begin{equation}
(\mathbf{K}+\delta \mathbf{K})(\boldsymbol{q}_1+\delta \boldsymbol{q}_1)
=
(\lambda_1+\delta\lambda_1)(\boldsymbol{q}_1+\delta \boldsymbol{q}_1).
\label{eq:supp_pert_eigen}
\end{equation}
Keeping first-order terms and enforcing the unit-norm constraint \(\boldsymbol{q}_1^{\mathrm{T}}\delta \boldsymbol{q}_1=0\) give
\begin{equation}
(\lambda_1\mathbf{I}_4-\mathbf{K})\delta \boldsymbol{q}_1
=
\left(\mathbf{I}_4-\boldsymbol{q}_1\boldsymbol{q}_1^{\mathrm{T}}\right)\delta \mathbf{K}\boldsymbol{q}_1 .
\label{eq:supp_dq_linear}
\end{equation}
Because \(\lambda_1\mathbf{I}_4-\mathbf{K}\) is singular along the \(\boldsymbol{q}_1\) direction, the rank-one term \(\boldsymbol{q}_1\boldsymbol{q}_1^{\mathrm{T}}\) is added to obtain a nonsingular system, yielding
\begin{equation}
\delta \boldsymbol{q}_1
=
\left(\lambda_1\mathbf{I}_4-\mathbf{K}+\boldsymbol{q}_1\boldsymbol{q}_1^{\mathrm{T}}\right)^{-1}
\left(\mathbf{I}_4-\boldsymbol{q}_1\boldsymbol{q}_1^{\mathrm{T}}\right)\delta \mathbf{K}\boldsymbol{q}_1 .
\label{eq:supp_dq}
\end{equation}
This modification does not affect \(\delta \boldsymbol{q}_1\) because the unit-norm constraint requires \(\boldsymbol{q}_1^{\mathrm{T}}\delta \boldsymbol{q}_1=0\).

\subsection{Quaternion Perturbation}
\label{sec:supp_quaternion}

Let \(\boldsymbol{q}_1=[s_1,\boldsymbol{\eta}_1^{\mathrm{T}}]^{\mathrm{T}}\) and \(\Delta \boldsymbol{q}_1=[\Delta s_1,\Delta\boldsymbol{\eta}_1^{\mathrm{T}}]^{\mathrm{T}}\). For scalar-first quaternion multiplication,
\begin{equation}
\boldsymbol{q}_1\odot\Delta \boldsymbol{q}_1
=
\begin{bmatrix}
s_1\Delta s_1-\boldsymbol{\eta}_1^{\mathrm{T}}\Delta\boldsymbol{\eta}_1\\
\Delta s_1\boldsymbol{\eta}_1+s_1\Delta\boldsymbol{\eta}_1
+\boldsymbol{\eta}_1\times\Delta\boldsymbol{\eta}_1
\end{bmatrix}.
\label{eq:supp_q_product}
\end{equation}
Under the adopted right-multiplicative convention, the perturbed quaternion is related to the nominal quaternion by
\begin{equation}
\boldsymbol{q}_1+\delta \boldsymbol{q}_1=\boldsymbol{q}_1\odot\Delta \boldsymbol{q}_1.
\label{eq:supp_mult_q}
\end{equation}
The exact axis--angle representation of the multiplicative error quaternion is
\begin{equation}
\Delta \boldsymbol{q}_1=
\begin{bmatrix}
\cos\!\left(\|\boldsymbol{\phi}^{i}\|/2\right)\\[0.4ex]
\dfrac{\boldsymbol{\phi}^{i}}{\|\boldsymbol{\phi}^{i}\|}
\sin\!\left(\|\boldsymbol{\phi}^{i}\|/2\right)
\end{bmatrix}.
\label{eq:supp_exact_mult_q}
\end{equation}
Using \(\cos(u/2)\simeq 1\) and \(\sin(u/2)/u\simeq 1/2\), the first-order small-misalignment approximation is
\begin{equation}
\Delta \boldsymbol{q}_1
\approx
\begin{bmatrix}
1\\
\tfrac{1}{2}\boldsymbol{\phi}^{i}
\end{bmatrix}.
\label{eq:supp_small_mult_q}
\end{equation}
Substituting \eqref{eq:supp_small_mult_q} into the quaternion product in \eqref{eq:supp_q_product} and subtracting \(\boldsymbol{q}_1\) give
\begin{equation}
\delta \boldsymbol{q}_1
\approx\frac{1}{2}\boldsymbol{\Xi}(\boldsymbol{q}_1)\boldsymbol{\phi}^{i},
\qquad
\boldsymbol{\Xi}(\boldsymbol{q}_1)=
\begin{bmatrix}
-\boldsymbol{\eta}_1^{\mathrm{T}}\\
s_1\mathbf{I}_3+\crossmat{\boldsymbol{\eta}_1}
\end{bmatrix}.
\label{eq:supp_dq_theta}
\end{equation}
Because \(\boldsymbol{q}_1\) is a unit quaternion, \(\boldsymbol{\Xi}(\boldsymbol{q}_1)^{\mathrm{T}}\boldsymbol{\Xi}(\boldsymbol{q}_1)=\mathbf{I}_3\). Thus, premultiplying \eqref{eq:supp_dq_theta} by \(2\boldsymbol{\Xi}(\boldsymbol{q}_1)^{\mathrm{T}}\) yields
\begin{equation}
\boldsymbol{\phi}^{i}
=2\boldsymbol{\Xi}(\boldsymbol{q}_1)^{\mathrm{T}}\delta \boldsymbol{q}_1,\qquad
\boldsymbol{\Xi}(\boldsymbol{q}_1)^{\mathrm{T}}
=\begin{bmatrix}
-\boldsymbol{\eta}_1 & s_1\mathbf{I}_3-\crossmat{\boldsymbol{\eta}_1}
\end{bmatrix}.
\label{eq:supp_Xi}
\end{equation}
\subsection{Misalignment-to-Euler-Angle Mapping}
\label{sec:supp_euler_mapping}

Under the ENU navigation-frame and RFU body-frame conventions, define the nominal and estimated Euler attitude angles as
\begin{equation}
\boldsymbol{\theta}
=\begin{bmatrix}\theta&\gamma&\psi\end{bmatrix}^{\mathrm{T}},
\qquad
\widetilde{\boldsymbol{\theta}}
=\begin{bmatrix}\widetilde{\theta}&\widetilde{\gamma}&\widetilde{\psi}\end{bmatrix}^{\mathrm{T}},
\label{eq:supp_euler_vectors}
\end{equation}
where \(\theta\), \(\gamma\), and \(\psi\) are pitch, roll, and heading, respectively. For a scalar-first quaternion \(\boldsymbol{q}=[s,\boldsymbol{\eta}^{\mathrm{T}}]^{\mathrm{T}}\), adopt the attitude-matrix definition
\begin{equation}
\begin{aligned}
\mathbf{C}(\boldsymbol{q})
&=\left(s^2-\boldsymbol{\eta}^{\mathrm{T}}\boldsymbol{\eta}\right)\mathbf{I}_3
+2\boldsymbol{\eta}\boldsymbol{\eta}^{\mathrm{T}}
-2s\crossmat{\boldsymbol{\eta}},\\
\mathbf{C}(\boldsymbol{q}_1)&=\mathbf{C}_{b(0)}^{n(0)}.
\end{aligned}
\label{eq:supp_Cq_definition}
\end{equation}
Using \(\mathbf{C}(\boldsymbol{q}_a\odot\boldsymbol{q}_b)=\mathbf{C}(\boldsymbol{q}_b)\mathbf{C}(\boldsymbol{q}_a)\), the right-multiplicative error in \eqref{eq:supp_mult_q} yields
\begin{equation}
\begin{aligned}
\widetilde{\mathbf{C}}_{b(0)}^{n(0)}
&=\mathbf{C}(\boldsymbol{q}_1\odot\Delta\boldsymbol{q}_1)
=\mathbf{C}(\Delta\boldsymbol{q}_1)\mathbf{C}(\boldsymbol{q}_1)\\
&\approx
\left(\mathbf{I}_3-\crossmat{\boldsymbol{\phi}^{i}}\right)
\mathbf{C}_{b(0)}^{n(0)}.
\end{aligned}
\label{eq:supp_right_mult_matrix}
\end{equation}
For the adopted pitch--roll--heading sequence, the first-order perturbation produced by the Euler attitude-angle error is
\begin{equation}
\widetilde{\mathbf{C}}_{b(0)}^{n(0)}
\approx
\mathbf{C}_{b(0)}^{n(0)}
\left(
\mathbf{I}_3+
\crossmat{\mathbf{C}_{A}^{\omega}\delta\boldsymbol{\theta}}
\right).
\label{eq:supp_euler_matrix_variation}
\end{equation}
Comparing \eqref{eq:supp_right_mult_matrix} and \eqref{eq:supp_euler_matrix_variation} yields
\begin{equation}
-\mathbf{C}_{n(0)}^{b(0)}\boldsymbol{\phi}^{i}
=\mathbf{C}_{A}^{\omega}\delta\boldsymbol{\theta},
\qquad
\delta\boldsymbol{\theta}
=\widetilde{\boldsymbol{\theta}}-\boldsymbol{\theta}.
\label{eq:supp_euler_local_relation}
\end{equation}
Here,
\begin{equation}
\mathbf{C}_{A}^{\omega}
=
\begin{bmatrix}
\cos\gamma & 0 & -\sin\gamma\cos\theta\\
0 & 1 & \sin\theta\\
\sin\gamma & 0 & \cos\gamma\cos\theta
\end{bmatrix}.
\label{eq:supp_CAomega}
\end{equation}
Solving \eqref{eq:supp_euler_local_relation} for the Euler attitude-angle error gives the first-order mapping
\begin{equation}
\delta\boldsymbol{\theta}
=\mathbf{J}_{\phi}\boldsymbol{\phi}^{i},
\qquad
\mathbf{J}_{\phi}
=-\left(\mathbf{C}_{A}^{\omega}\right)^{-1}
\mathbf{C}_{n(0)}^{b(0)}.
\label{eq:supp_Jphi_definition}
\end{equation}
The explicit Jacobian is
\begin{equation}
\mathbf{J}_{\phi}
=-
\begin{bmatrix}
\cos\gamma & 0 & \sin\gamma\\
\sin\gamma\tan\theta & 1 & -\cos\gamma\tan\theta\\
-\dfrac{\sin\gamma}{\cos\theta} & 0 & \dfrac{\cos\gamma}{\cos\theta}
\end{bmatrix}
\mathbf{C}_{n(0)}^{b(0)}.
\label{eq:supp_Jphi}
\end{equation}
Combining \eqref{eq:supp_dq}, \eqref{eq:supp_Xi}, and \eqref{eq:supp_Jphi_definition} yields the Euler attitude-error mapping used in the main paper,
\begin{equation}
\delta\boldsymbol{\theta}=\boldsymbol{\Phi}\operatorname{vec}(\delta \mathbf{K}).
\label{eq:supp_theta}
\end{equation}
\subsection{Observation-Vector Jacobians}
\label{sec:supp_jacobians}

This section gives the explicit forms of the Jacobian matrices used in the main paper. Because the main derivation uses the Davenport gain matrix \(\mathbf{K}\) in the maximization form, the Jacobians differ from those obtained from the symmetric quadratic-form matrix \(\mathbf{K}_{\mathrm{OBA}}\). Direct differentiation gives \(\delta \mathbf{K}_{\mathrm{OBA}}=\delta c\,\mathbf{I}_4-2\delta \mathbf{K}\); therefore, its Jacobians contain a factor of \(-2\) and an additional identity-matrix term \(\delta c\,\mathbf{I}_4\).

For a single observation pair, let
\begin{equation}
\boldsymbol{\beta}_i=[\beta_{i1},\beta_{i2},\beta_{i3}]^{\mathrm{T}},
\qquad
\boldsymbol{\alpha}_i=[\alpha_{i1},\alpha_{i2},\alpha_{i3}]^{\mathrm{T}}.
\label{eq:supp_components}
\end{equation}
Using \(\operatorname{tr}(\boldsymbol{a}\boldsymbol{b}^{\mathrm{T}})=\boldsymbol{b}^{\mathrm{T}}\boldsymbol{a}\), the first-order perturbations entering \(\mathbf{K}\) are
\begin{align}
\delta \mathbf{B}_i&=w_i
\left(
\delta\boldsymbol{\beta}_i\boldsymbol{\alpha}_i^{\mathrm{T}}
+\boldsymbol{\beta}_i\delta\boldsymbol{\alpha}_i^{\mathrm{T}}
\right),
\label{eq:supp_dB}\\[-0.4ex]
\operatorname{tr}(\delta \mathbf{B}_i)&=w_i
\left(
\boldsymbol{\alpha}_i^{\mathrm{T}}\delta\boldsymbol{\beta}_i
+\boldsymbol{\beta}_i^{\mathrm{T}}\delta\boldsymbol{\alpha}_i
\right).
\label{eq:supp_trdB}
\end{align}
\begin{equation}
\delta\boldsymbol{z}_i=w_i
\left(
\delta\boldsymbol{\beta}_i\times\boldsymbol{\alpha}_i
+\boldsymbol{\beta}_i\times\delta\boldsymbol{\alpha}_i
\right).
\label{eq:supp_dZ}
\end{equation}
Substitution of \eqref{eq:supp_dB}--\eqref{eq:supp_dZ} into the Davenport gain matrix perturbation and column-wise vectorization give
\begin{equation}
\operatorname{vec}(\delta \mathbf{K}_i)
=\mathbf{J}_{\beta,i}\delta\boldsymbol{\beta}_i
+\mathbf{J}_{\alpha,i}\delta\boldsymbol{\alpha}_i,
\label{eq:supp_J_relation}
\end{equation}
where \(\mathbf{J}_{\beta,i},\mathbf{J}_{\alpha,i}\in\mathbb{R}^{16\times3}\) are obtained by differentiating each element of \(\operatorname{vec}(\mathbf{K})\) with respect to the components of \(\boldsymbol{\beta}_i\) and \(\boldsymbol{\alpha}_i\). Under the column-major ordering used by \(\operatorname{vec}(\cdot)\),
\begin{equation}
\mathbf{J}_{\beta,i}
=
w_i
\begin{bmatrix}
\alpha_{i1} & \alpha_{i2} & \alpha_{i3} \\
0 & \alpha_{i3} & -\alpha_{i2} \\
-\alpha_{i3} & 0 & \alpha_{i1} \\
\alpha_{i2} & -\alpha_{i1} & 0 \\
0 & \alpha_{i3} & -\alpha_{i2} \\
\alpha_{i1} & -\alpha_{i2} & -\alpha_{i3} \\
\alpha_{i2} & \alpha_{i1} & 0 \\
\alpha_{i3} & 0 & \alpha_{i1} \\
-\alpha_{i3} & 0 & \alpha_{i1} \\
\alpha_{i2} & \alpha_{i1} & 0 \\
-\alpha_{i1} & \alpha_{i2} & -\alpha_{i3} \\
0 & \alpha_{i3} & \alpha_{i2} \\
\alpha_{i2} & -\alpha_{i1} & 0 \\
\alpha_{i3} & 0 & \alpha_{i1} \\
0 & \alpha_{i3} & \alpha_{i2} \\
-\alpha_{i1} & -\alpha_{i2} & \alpha_{i3}
\end{bmatrix}.
\label{eq:supp_Jbeta}
\end{equation}
\begin{equation}
\mathbf{J}_{\alpha,i}
=
w_i
\begin{bmatrix}
\beta_{i1} & \beta_{i2} & \beta_{i3} \\
0 & -\beta_{i3} & \beta_{i2} \\
\beta_{i3} & 0 & -\beta_{i1} \\
-\beta_{i2} & \beta_{i1} & 0 \\
0 & -\beta_{i3} & \beta_{i2} \\
\beta_{i1} & -\beta_{i2} & -\beta_{i3} \\
\beta_{i2} & \beta_{i1} & 0 \\
\beta_{i3} & 0 & \beta_{i1} \\
\beta_{i3} & 0 & -\beta_{i1} \\
\beta_{i2} & \beta_{i1} & 0 \\
-\beta_{i1} & \beta_{i2} & -\beta_{i3} \\
0 & \beta_{i3} & \beta_{i2} \\
-\beta_{i2} & \beta_{i1} & 0 \\
\beta_{i3} & 0 & \beta_{i1} \\
0 & \beta_{i3} & \beta_{i2} \\
-\beta_{i1} & -\beta_{i2} & \beta_{i3}
\end{bmatrix}.
\label{eq:supp_Jalpha}
\end{equation}
Stacking all observation pairs yields the global Jacobian \(\mathbf{J}_{\mathrm{K}}\) used in the main paper and hence the complete first-order attitude error mapping.

\section{Discrete Observation-Error Propagation and Global-Epoch Assembly}
\label{sec:supp_global}

This section discretizes the continuous observation-vector perturbations derived in Sec.~V into single-step contributions and sliding-window sums. The resulting discrete model is then used to derive the deterministic attitude offset and the covariance of the stochastic attitude error.

\subsection{Discrete Projection-Error Recursions}
\label{sec:supp_projection_discrete}

From the continuous projection-error propagation in \eqref{eq:supp_phi}, the body-frame case is obtained by setting \(x=b\). Separating the accumulated error at \(t_k\) and the newly injected gyroscope error over \([t_k,t_{k+1}]\), the discrete recursion becomes
\begin{equation}
\boldsymbol{\phi}^b(t_{k+1})
\approx
\mathbf{C}_{b(t_k)}^{b(t_{k+1})}
\boldsymbol{\phi}^b(t_k)
+T\delta\boldsymbol{\omega}_{ib,k}^b.
\label{eq:supp_phi_body_disc}
\end{equation}
For the navigation frame, the corresponding recursion is
\begin{equation}
\boldsymbol{\phi}^n(t_{k+1})
\approx
\mathbf{C}_{n(t_k)}^{n(t_{k+1})}
\boldsymbol{\phi}^n(t_k)
+T\mathbf{M}_{v,k}
\delta\boldsymbol{v}_{{\rm GNSS},k}^n.
\label{eq:supp_phi_nav_disc}
\end{equation}
Thus, \(\boldsymbol{\phi}^b\) is driven directly by gyroscope error, whereas \(\boldsymbol{\phi}^n\) is driven by GNSS velocity error through \(\mathbf{M}_v\). The latter mapping is strongly attenuated by the Earth-curvature radii contained in \(\mathbf{M}_v\).

\subsection{Body-Side Single-Step Discretization}
\label{sec:supp_body_discrete}

The single-step contributions over the \(k\)th sampling interval are
\begin{equation}
\begin{aligned}
\Delta\delta\boldsymbol{\alpha}_{a,k}
&=\int_{t_k}^{t_{k+1}}
\mathbf{C}_{b(\tau)}^{b(0)}
\delta\boldsymbol{f}^b(\tau)d\tau,\\
\Delta\delta\boldsymbol{\alpha}_{g,k}
&=-\int_{t_k}^{t_{k+1}}
\mathbf{C}_{b(\tau)}^{b(0)}
\crossmat{\boldsymbol{f}^b(\tau)}
\boldsymbol{\phi}^b(\tau)d\tau .
\end{aligned}
\label{eq:supp_alpha_step_exact}
\end{equation}
Factoring out the attitude-chain matrix at the interval start gives
\begin{equation}
\begin{aligned}
\Delta\delta\boldsymbol{\alpha}_{a,k}
&=\mathbf{C}_{b(t_k)}^{b(0)}
\int_{t_k}^{t_{k+1}}
\mathbf{C}_{b(\tau)}^{b(t_k)}
\delta\boldsymbol{f}^b(\tau)d\tau,\\
\Delta\delta\boldsymbol{\alpha}_{g,k}
&=-\mathbf{C}_{b(t_k)}^{b(0)}
\int_{t_k}^{t_{k+1}}
\mathbf{C}_{b(\tau)}^{b(t_k)}
\crossmat{\boldsymbol{f}^b(\tau)}
\boldsymbol{\phi}^b(\tau)d\tau .
\end{aligned}
\label{eq:supp_alpha_step_factored}
\end{equation}
Approximating \(\delta\boldsymbol{f}^b\) and \(\boldsymbol{\phi}^b\) as constant within one sampling interval yields
\begin{equation}
\begin{aligned}
\Delta\delta\boldsymbol{\alpha}_{a,k}
&\approx\mathbf{C}_{b(t_k)}^{b(0)}
\left[
\int_{t_k}^{t_{k+1}}
\mathbf{C}_{b(\tau)}^{b(t_k)}d\tau
\right]
\delta\boldsymbol{f}_k^b,\\
\Delta\delta\boldsymbol{\alpha}_{g,k}
&\approx-\mathbf{C}_{b(t_k)}^{b(0)}
\crossmat{\Delta\boldsymbol{v}_{c,k}}
\boldsymbol{\phi}^b(t_k).
\end{aligned}
\label{eq:supp_alpha_step_approx}
\end{equation}
Applying the dual-sample rotation and sculling result \eqref{eq:supp_dvc} to the two half-period accelerometer velocity-increment errors gives
\begin{equation}
\begin{aligned}
\mathbf{A}_{a,1,k}&=\mathbf{C}_{b(t_k)}^{b(0)}
\Bigl\{\mathbf{I}_3+\tfrac12\crossmat{\Delta\boldsymbol{\theta}_{1,k}+\Delta\boldsymbol{\theta}_{2,k}}\\
&\qquad-\tfrac23\crossmat{\Delta\boldsymbol{\theta}_{2,k}}\Bigr\},\\
\mathbf{A}_{a,2,k}&=\mathbf{C}_{b(t_k)}^{b(0)}
\Bigl\{\mathbf{I}_3+\tfrac12\crossmat{\Delta\boldsymbol{\theta}_{1,k}+\Delta\boldsymbol{\theta}_{2,k}}\\
&\qquad+\tfrac23\crossmat{\Delta\boldsymbol{\theta}_{1,k}}\Bigr\},\\
\mathbf{A}_{a,k}&=\begin{bmatrix}\mathbf{A}_{a,1,k}&\mathbf{A}_{a,2,k}\end{bmatrix}
\in\mathbb{R}^{3\times6},\\
\boldsymbol{w}_{a,k}&=\begin{bmatrix}\delta\boldsymbol{v}_{a,1,k}\\
\delta\boldsymbol{v}_{a,2,k}\end{bmatrix}\in\mathbb{R}^{6},\\
\Delta\delta\boldsymbol{\alpha}_{a,k}
&=\mathbf{A}_{a,1,k}\delta\boldsymbol{v}_{a,1,k}
+\mathbf{A}_{a,2,k}\delta\boldsymbol{v}_{a,2,k}\\
&=\mathbf{A}_{a,k}\boldsymbol{w}_{a,k}.
\end{aligned}
\label{eq:supp_Aak}
\end{equation}
The two half-period velocity-increment errors are generally distinct inputs; they become equal only when the model is specialized to a constant accelerometer bias. Summing the single-step contributions over a sliding window gives
\begin{equation}
\delta\boldsymbol{\alpha}_{a,i}
=\sum_{k=m_i}^{M_i-1}
\Delta\delta\boldsymbol{\alpha}_{a,k},
\qquad
\delta\boldsymbol{\alpha}_{g,i}
=\sum_{k=m_i}^{M_i-1}
\Delta\delta\boldsymbol{\alpha}_{g,k}.
\label{eq:supp_alpha_window_sum}
\end{equation}

\subsection{Navigation-Side Single-Step Discretization}
\label{sec:supp_navigation_discrete}

Using the navigation-frame projection-error recursion \eqref{eq:supp_phi_nav_disc}, the integral contribution of the GNSS velocity error over one sampling interval is
\begin{equation}
\begin{aligned}
\Delta\delta\boldsymbol{\beta}_{v,k}
={}&\int_{t_k}^{t_{k+1}}
\mathbf{C}_{n(\tau)}^{n(0)}
\Bigl\{
\boldsymbol{\omega}_{ie}^n(\tau)\times
\delta\boldsymbol{v}_{\rm GNSS}^n(\tau)\\
&\quad+\boldsymbol{\phi}^n(\tau)\times
\left[
\boldsymbol{\omega}_{ie}^n(\tau)\times\boldsymbol{v}^n(\tau)
-\boldsymbol{g}^n(\tau)
\right]
\Bigr\}d\tau .
\end{aligned}
\label{eq:supp_beta_v_step}
\end{equation}
Therefore, the complete discrete sliding-window GNSS velocity error is
\begin{equation}
\begin{aligned}
\delta\boldsymbol{\beta}_{v,i}
={}&\mathbf{C}_{n(t_{M_i})}^{n(0)}
\delta\boldsymbol{v}_{\rm GNSS}^n(t_{M_i})
-\mathbf{C}_{n(t_{m_i})}^{n(0)}
\delta\boldsymbol{v}_{\rm GNSS}^n(t_{m_i})\\
&+\mathbf{C}_{n(t_{M_i})}^{n(0)}
\crossmat{\boldsymbol{\phi}^n(t_{M_i})}
\boldsymbol{v}^n(t_{M_i})\\
&-\mathbf{C}_{n(t_{m_i})}^{n(0)}
\crossmat{\boldsymbol{\phi}^n(t_{m_i})}
\boldsymbol{v}^n(t_{m_i})\\
&+\sum_{k=m_i}^{M_i-1}
\Delta\delta\boldsymbol{\beta}_{v,k}.
\end{aligned}
\label{eq:supp_beta_v_window_full}
\end{equation}
The direct Earth-rate integral is smaller than the endpoint term by a factor of order \(\|\boldsymbol{\omega}_{ie}^n\|T_w\), approximately \(7.3\times10^{-4}\) for \(T_w=10\) s. The navigation-frame projection error is additionally attenuated by the curvature-dependent factors \(T/R_E\) and \(vT/R_E\). With \(T\leq0.02\) s, \(R_E\simeq6.37\times10^6\) m, and \(v\leq20\) m/s, these factors are below \(3.2\times10^{-9}\) s/m and \(6.3\times10^{-8}\), respectively. Consequently, both the \(\boldsymbol{\phi}^n\)-induced endpoint terms and the accumulated single-step integral terms are much smaller than the direct GNSS velocity endpoint contribution, giving
\begin{equation}
\delta\boldsymbol{\beta}_{v,i}
\approx
\mathbf{C}_{n(t_{M_i})}^{n(0)}
\delta\boldsymbol{v}_{\rm GNSS}^n(t_{M_i})
-\mathbf{C}_{n(t_{m_i})}^{n(0)}
\delta\boldsymbol{v}_{\rm GNSS}^n(t_{m_i}).
\label{eq:supp_beta_v_endpoint}
\end{equation}

The Earth-rate coupling in the deterministic lever-arm term contributes over one sampling interval as
\begin{equation}
\Delta\delta\boldsymbol{\alpha}_{l,k}
=\int_{t_k}^{t_{k+1}}
\mathbf{C}_{b(\tau)}^{b(0)}
\left\{
\boldsymbol{\omega}_{ie}^b(\tau)\times
\left[
\boldsymbol{\omega}_{eb}^b(\tau)\times\boldsymbol{l}^b
\right]
\right\}d\tau .
\label{eq:supp_alpha_l_step}
\end{equation}
The complete discrete sliding-window lever-arm contribution is
\begin{equation}
\begin{aligned}
\delta\boldsymbol{\alpha}_{l,i}
={}&\mathbf{C}_{b(t_{M_i})}^{b(0)}
\left[
\boldsymbol{\omega}_{eb}^b(t_{M_i})\times\boldsymbol{l}^b
\right]\\
&-\mathbf{C}_{b(t_{m_i})}^{b(0)}
\left[
\boldsymbol{\omega}_{eb}^b(t_{m_i})\times\boldsymbol{l}^b
\right]
+\sum_{k=m_i}^{M_i-1}
\Delta\delta\boldsymbol{\alpha}_{l,k}.
\end{aligned}
\label{eq:supp_alpha_l_window_full}
\end{equation}
Since \(\boldsymbol{\omega}_{eb}^b=\boldsymbol{\omega}_{ib}^b-\boldsymbol{\omega}_{ie}^b\approx\boldsymbol{\omega}_{ib}^b\), the accumulated Earth-rate coupling is bounded by
\begin{equation}
\left\|
\sum_{k=m_i}^{M_i-1}
\Delta\delta\boldsymbol{\alpha}_{l,k}
\right\|
\leq\|\boldsymbol{\omega}_{ie}^b\|T_w
\max_{\tau\in[t_{m_i},t_{M_i}]}
\left\|\boldsymbol{\omega}_{eb}^b(\tau)\times\boldsymbol{l}^b\right\|.
\label{eq:supp_lever_bound}
\end{equation}
It is smaller than the retained endpoint term by a factor of order \(\|\boldsymbol{\omega}_{ie}^b\|T_w\). The retained lever-arm model is therefore
\begin{equation}
\begin{aligned}
\delta\boldsymbol{\alpha}_{l,i}
\approx{}&\mathbf{C}_{b(t_{M_i})}^{b(0)}
\left[
\boldsymbol{\omega}_{ib}^b(t_{M_i})\times\boldsymbol{l}^b
\right]\\
&-\mathbf{C}_{b(t_{m_i})}^{b(0)}
\left[
\boldsymbol{\omega}_{ib}^b(t_{m_i})\times\boldsymbol{l}^b
\right].
\end{aligned}
\label{eq:supp_alpha_l_endpoint}
\end{equation}

\subsection{Deterministic Attitude-Offset Computation}
\label{sec:supp_deterministic}

For a constant accelerometer bias \(\boldsymbol{\nabla}^b\), the two half-period velocity-increment errors in \eqref{eq:supp_Aak} satisfy
\begin{equation}
\begin{aligned}
\delta\boldsymbol{v}_{a,1,k}
&=\delta\boldsymbol{v}_{a,2,k}
=\frac{T}{2}\boldsymbol{\nabla}^b,\\
\delta\boldsymbol{\alpha}_{a,i}(\boldsymbol{\nabla}^b)
&=\sum_{k=m_i}^{M_i-1}\mathbf{A}_{a,k}
\begin{bmatrix}
\dfrac{T}{2}\boldsymbol{\nabla}^b\\[1mm]
\dfrac{T}{2}\boldsymbol{\nabla}^b
\end{bmatrix}.
\end{aligned}
\label{eq:supp_alpha_a_det}
\end{equation}

For a constant gyroscope bias \(\boldsymbol{\epsilon}^b\), setting \(\delta\boldsymbol{\omega}_{ib,k}^b=\boldsymbol{\epsilon}^b\) in \eqref{eq:supp_phi_body_disc} gives
\begin{equation}
\begin{aligned}
\boldsymbol{\phi}_{\epsilon}^b(t_{k+1})
&\approx\mathbf{C}_{b(t_k)}^{b(t_{k+1})}\boldsymbol{\phi}_{\epsilon}^b(t_k)
+T\boldsymbol{\epsilon}^b,\\
\boldsymbol{\phi}_{\epsilon}^b(t_0)&=\boldsymbol{0}.
\end{aligned}
\label{eq:supp_phi_eps}
\end{equation}
The resulting contribution to the body-side observation-vector error over the \(i\)th sliding window is
\begin{equation}
\delta\boldsymbol{\alpha}_{g,i}(\boldsymbol{\epsilon}^b)
=-\sum_{j=m_i}^{M_i-1}\mathbf{C}_{b(t_j)}^{b(0)}
\crossmat{\Delta\boldsymbol{v}_{c,j}}
\boldsymbol{\phi}_{\epsilon}^b(t_j).
\label{eq:supp_alpha_g_det}
\end{equation}

Together with the lever-arm endpoint model \eqref{eq:supp_alpha_l_endpoint}, these deterministic inputs give
\begin{equation}
\delta\boldsymbol{\theta}_s
=\boldsymbol{\Phi}\sum_{i=1}^{N}\mathbf{J}_{\alpha,i}
\left[
\delta\boldsymbol{\alpha}_{g,i}(\boldsymbol{\epsilon}^b)
+\delta\boldsymbol{\alpha}_{a,i}(\boldsymbol{\nabla}^b)
-\delta\boldsymbol{\alpha}_{l,i}(\boldsymbol{l}^b)
\right].
\label{eq:supp_theta_s}
\end{equation}

\subsection{Stochastic Attitude-Error Covariance Computation}
\label{sec:supp_covariance}

Let \(\boldsymbol{w}_{a,k}\in\mathbb{R}^6\) denote the stacked pair of half-period accelerometer velocity-increment noise vectors defined in \eqref{eq:supp_Aak}, and let \(\boldsymbol{w}_{g,k},\boldsymbol{w}_{v,k}\in\mathbb{R}^3\) denote the gyroscope and GNSS velocity-noise increments, respectively. For the \(i\)th sliding window, the stochastic contributions of the accelerometer noise, gyroscope noise, and GNSS velocity noise to the observation vectors are
\begin{equation}
\begin{aligned}
\delta\boldsymbol{\alpha}_{a,i}
&=\sum_{k=m_i}^{M_i-1}\mathbf{A}_{a,k}\boldsymbol{w}_{a,k},\\
\delta\boldsymbol{\alpha}_{g,i}
&=-\sum_{j=m_i}^{M_i-1}\mathbf{C}_{b(t_j)}^{b(0)}
\crossmat{\Delta\boldsymbol{v}_{c,j}}
\sum_{k=0}^{j-1}\mathbf{C}_{b(t_{k+1})}^{b(t_j)}T\boldsymbol{w}_{g,k},\\
\delta\boldsymbol{\beta}_{v,i}
&=\mathbf{C}_{n(t_{M_i})}^{n(0)}\boldsymbol{w}_{v,M_i}
-\mathbf{C}_{n(t_{m_i})}^{n(0)}\boldsymbol{w}_{v,m_i}.
\end{aligned}
\label{eq:supp_window_random}
\end{equation}
Substituting the single-window contributions into the attitude-error mapping and exchanging the window and global-epoch sums give
\begin{equation}
\delta\boldsymbol{\theta}_r
=\boldsymbol{\Phi}\sum_{k=1}^{K_{\max}}
\left(
\mathbf{X}_k\boldsymbol{w}_{a,k}
+\mathbf{Y}_k\boldsymbol{w}_{g,k}
+\mathbf{Z}_k\boldsymbol{w}_{v,k}
\right),
\label{eq:supp_epoch}
\end{equation}
where
\begin{align}
\mathbf{X}_k
&=\sum_{i\in\Omega_k}\mathbf{J}_{\alpha,i}\mathbf{A}_{a,k},
\label{eq:supp_Xk}\\
\mathbf{Y}_k
&=-\sum_{i=1}^{N}\mathbf{J}_{\alpha,i}
\sum_{j=\max(m_i,k+1)}^{M_i-1}
\mathbf{C}_{b(t_j)}^{b(0)}\crossmat{\Delta\boldsymbol{v}_{c,j}}
\mathbf{C}_{b(t_{k+1})}^{b(t_j)}T,
\label{eq:supp_Yk}\\
\mathbf{Z}_k
&=\sum_{i\in\Omega_k^{\mathrm{end}}}\mathbf{J}_{\beta,i}\mathbf{C}_{n(t_k)}^{n(0)}
-\sum_{i\in\Omega_k^{\mathrm{start}}}\mathbf{J}_{\beta,i}\mathbf{C}_{n(t_k)}^{n(0)}.
\label{eq:supp_Zk}
\end{align}
Here, \(\mathbf{A}_{a,k}\) maps the two half-period accelerometer velocity-increment noise vectors in \(\boldsymbol{w}_{a,k}\) to the body-side observation-vector perturbation. The set \(\Omega_k=\{i\mid m_i\leq k\leq M_i-1\}\) identifies the sliding windows covering the \(k\)th specific-force integration interval, and \(\Delta\boldsymbol{v}_{c,j}\) is the compensated specific-force increment defined in Sec.~II. Because \(\boldsymbol{w}_{g,k}\) acts over \([t_k,t_{k+1}]\), it first contributes to the projection error at \(t_{k+1}\), which gives the lower limit \(j=\max(m_i,k+1)\). If this index range is empty, the inner sum is zero. The endpoint sets \(\Omega_k^{\mathrm{start}}=\{i\mid m_i=k\}\) and \(\Omega_k^{\mathrm{end}}=\{i\mid M_i=k\}\) identify the sliding windows starting and ending at \(t_k\), respectively.

Equation \eqref{eq:supp_Xk} accounts for accelerometer samples shared by overlapping windows, \eqref{eq:supp_Yk} propagates gyroscope noise through the accumulated body-attitude projection error, and \eqref{eq:supp_Zk} aggregates the GNSS velocity-noise contributions from all sliding windows having epoch \(t_k\) as an endpoint.

The input covariance matrices are defined by
\begin{equation}
\begin{aligned}
\mathbf{Q}_{a,k}&=\operatorname{Cov}(\boldsymbol{w}_{a,k}),\\
\mathbf{Q}_{g,k}&=\operatorname{Cov}(\boldsymbol{w}_{g,k}),\\
\mathbf{R}_{v,k}&=\operatorname{Cov}(\boldsymbol{w}_{v,k}).
\end{aligned}
\label{eq:supp_input_covariances}
\end{equation}
For independent noise increments over the two half periods,
\begin{equation}
\mathbf{Q}_{a,k}
=\operatorname{blkdiag}\left(\mathbf{Q}_{a,1,k},\mathbf{Q}_{a,2,k}\right)
\in\mathbb{R}^{6\times6},
\label{eq:supp_accel_noise_covariance}
\end{equation}
whereas \(\mathbf{Q}_{g,k},\mathbf{R}_{v,k}\in\mathbb{R}^{3\times3}\). Under the mutually independent white-noise assumption, the global-epoch covariance is
\begin{equation}
\begin{aligned}
\mathbf{P}_{\theta,r}
&=\boldsymbol{\Phi}\Biggl[
\sum_{k=1}^{K_{\max}}
\Bigl(
\mathbf{X}_k\mathbf{Q}_{a,k}\mathbf{X}_k^{\mathrm{T}}
+\mathbf{Y}_k\mathbf{Q}_{g,k}\mathbf{Y}_k^{\mathrm{T}}\\
&\qquad
+\mathbf{Z}_k\mathbf{R}_{v,k}\mathbf{Z}_k^{\mathrm{T}}
\Bigr)\Biggr]\boldsymbol{\Phi}^{\mathrm{T}}.
\end{aligned}
\label{eq:supp_cov}
\end{equation}
For isotropic GNSS velocity noise with an epoch-independent variance, \(\mathbf{R}_{v,k}=\sigma_v^2\mathbf{I}_3\).

Let
\[
\delta\boldsymbol{y}_i
=
\begin{bmatrix}
\delta\boldsymbol{\beta}_i^{\mathrm{T}} &
\delta\boldsymbol{\alpha}_i^{\mathrm{T}}
\end{bmatrix}^{\mathrm{T}},
\qquad
\mathbf{P}_{y,ij}
=\operatorname{E}\left[
\delta\boldsymbol{y}_i\delta\boldsymbol{y}_j^{\mathrm{T}}
\right],
\]
where \(\mathbf{P}_{y,ij}\) is the cross-covariance of the \(i\)th and \(j\)th sliding-window observation errors. With \(\mathbf{H}_i=[\mathbf{J}_{\beta,i}\ \mathbf{J}_{\alpha,i}]\), propagation through the attitude-error mapping gives
\begin{equation}
\mathbf{P}_{\theta,r}=\boldsymbol{\Phi}
\left(\sum_{i=1}^{N}\sum_{j=1}^{N}\mathbf{H}_i\mathbf{P}_{y,ij}\mathbf{H}_j^{\mathrm{T}}\right)
\boldsymbol{\Phi}^{\mathrm{T}}.
\label{eq:supp_window_cov}
\end{equation}
Equation~\eqref{eq:supp_window_cov} explicitly retains the cross-covariances caused by overlapping sliding windows. Regrouping all contributions associated with each shared noise sample incorporates these cross-window correlations into the aggregated coefficients \(\mathbf{X}_k\), \(\mathbf{Y}_k\), and \(\mathbf{Z}_k\). Since noise samples from distinct epochs and sensor sources are assumed independent, the resulting covariance is equivalently expressed by \eqref{eq:supp_cov}.

\end{document}